\RequirePackage{fix-cm} 

\documentclass[epjc3,english,pdftex,twocolumn]{svjour3}

\usepackage{babel}
\usepackage{flushend}
\usepackage[T1]{fontenc}
\usepackage{graphicx}
\usepackage[switch]{lineno}
\usepackage{mathptmx} 
\usepackage[numbers,sort&compress]{natbib}
\usepackage{textcomp}
\usepackage{upgreek}
\usepackage{xcolor}
\usepackage{booktabs}
\usepackage{multirow}
\usepackage{threeparttable}
\usepackage{colortbl}
\usepackage{dcolumn}
\usepackage{microtype}

\usepackage{amsmath}
\usepackage{amssymb}

\usepackage[colorlinks,citecolor=blue,urlcolor=blue,linkcolor=blue]{hyperref}

\journalname{Eur. Phys. J. C}
\smartqed  

\begin{document}

\title{Characterization of alpha and beta interactions in liquid xenon}

\author{Florian~J\"org\thanksref{e3,addr1}
        \and
        Dominick~Cichon\thanksref{e1,addr1}
        \and
        Guillaume~Eurin\thanksref{e2,e7,addr1}
        \and
        Luisa H\"otzsch\thanksref{addr1}
        \and
        Teresa~Marrod\'an~Undagoitia\thanksref{e5,addr1}
        \and
        Natascha~Rupp\thanksref{e6,addr1}
}

\thankstext{e3}{e-mail: florian.joerg@mpi-hd.mpg.de}
\thankstext{e1}{e-mail: dominick.cichon@mpi-hd.mpg.de}
\thankstext{e2}{e-mail: guillaume.eurin@cea.fr}
\thankstext{e5}{e-mail: teresa.marrodan@mpi-hd.mpg.de}
\thankstext{e6}{e-mail: natascha.rupp@mpi-hd.mpg.de}
\thankstext{e7}{\emph{Present Address:} IRFU, CEA, Universit\'e Paris-Saclay, F-91191 Gif-sur-Yvette, France }

\institute{Max-Planck-Institut f\"ur Kernphysik, Saupfercheckweg 1, 69117 Heidelberg,
Germany\label{addr1}
}

\date{Received: date / Accepted: date}

\maketitle

\begin{abstract}
Liquid xenon based detectors have achieved great sensitivities in rare event searches.
Precise knowledge of the scintillation and ionization responses of the medium is essential to correctly model different interaction types in the detector including both signal and background-like ones.
The response of liquid xenon to low energy electrons and to alpha particles has been studied in the Heidelberg Xenon (HeXe) dual-phase xenon TPC.
We determine the light and charge signal yields for keV-energy electrons and MeV-energy alpha particles as well as the electron drift velocity for electric drift fields between  7.5 and 1\,640\,V/cm. A three dimensional simulation using COMSOL Multi\-physics\textsuperscript{\tiny\textregistered} is used to characterize the applied drift field and its homogeneity.
\end{abstract}

	
\section{Introduction}
\label{sec:intro}
	
Over the past decades, xenon time projection chambers (TPCs) have become one of the leading detector technologies for rare event searches, such as the search for neutrinoless double-$\upbeta$ decay\,\cite{Albert:2017owj} or for the elastic scattering of dark matter off target nuclei\,\cite{Undagoitia:2015gya}. Key advantages of this technology are the low energy threshold of $\mathcal{O}(\textrm{keV})$, the ultra low background and the large masses that can be realized.
The next generation of detectors is being commissioned or planned with masses at the multi-ton scale: XENONnT\,\cite{Aprile:2020vtw}, LZ\,\cite{LZ:2019sgr}, PandaX-4T\,\cite{PandaX:2018wtu}, nEXO\,\cite{Albert:2017hjq} or DARWIN\,\cite{Aalbers:2016jon}. 
\vskip 0.03cm

In dual-phase liquid xenon TPCs, particle interactions within the active detector volume can be detected via signals originating from excitation and ionization of the medium. Prompt scintillation light (S1) is emitted in the de-excitation of xenon excimer states created when the medium is excited by a particle's energy deposition. Alongside the scintillation, xenon atoms can also become ionized. Free electrons produced at the interaction site can be collected if an electric drift field is applied. The electrons drift vertically and are extracted to a gas phase on top of the liquid. In the gas phase a strong electric field amplifies the signal (S2) via proportional scintillation\,\cite{Lansiart:1976}. From the time difference between the S1 and S2 signals, the depth of the interaction can be determined. For detectors with several photosensors on top and bottom of the target, the photon hit distribution on the sensors can be used to reconstruct the $(x,y)$-coordinates of each event. Combined with the depth measurement, this provides a full 3-dimensional reconstruction of the event's position.
\vskip 0.02cm

The Heidelberg Xenon (HeXe) TPC has been developed as an R\&D instrument towards improving the understanding of various aspects relevant for current experiments searching for rare events. 
The system has been previously employed to test the response of R11410 PMTs of XENON1T to xenon scintillation light\,\cite{Barrow:2016doe}, to test radon reduction via boil-off distillation\,\cite{Bruenner:2016ziq}, and to test the impact of PTFE cleaning recipes for radon daughter removal on TPC performance\,\cite{Bruenner:2020arp}. In this manuscript, we detail the TPC setup and report on the response of liquid xenon to alpha (MeV energies from $^{222}$Rn, $^{218}$Po \& $^{214}$Po) and electron (keV energies from ${}^\mathrm{83m}\mathrm{Kr}$) interactions as a function of electric field. Note that only little literature is available for interactions of alpha particles\,\cite{Aprile:1991xb, Aprile:2006kx}. We investigate the light and charge yield dependence of alpha-induced energy depositions for fields between  7.5 and 1\,640\,V/cm. A detailed 3D electric-field model based on the finite element method (FEM) is used to precisely determine the electric field, its homogeneity and uncertainty. 

In section\,\ref{sec:hexe}, we describe the HeXe TPC setup and in section\,\ref{sec:efield} the electric field simulations performed in order to obtain a uniform electric field. Afterwards, section\,\ref{sec:systems}  describes the auxiliary systems and section\,\ref{sec:DAQ_processing_analysis} the data flow from data acquisition to data analysis.
Sections\,\ref{sec:response} and~\ref{sec:drift} contain the results including the measurements of charge yields, light yields for electrons and alpha particles, as well as a drift velocity determination for various electric fields.

	
\section{The HeXe detector}
\label{sec:hexe}

\begin{figure}[h]
		\centering
		\includegraphics[width=0.4\textwidth]{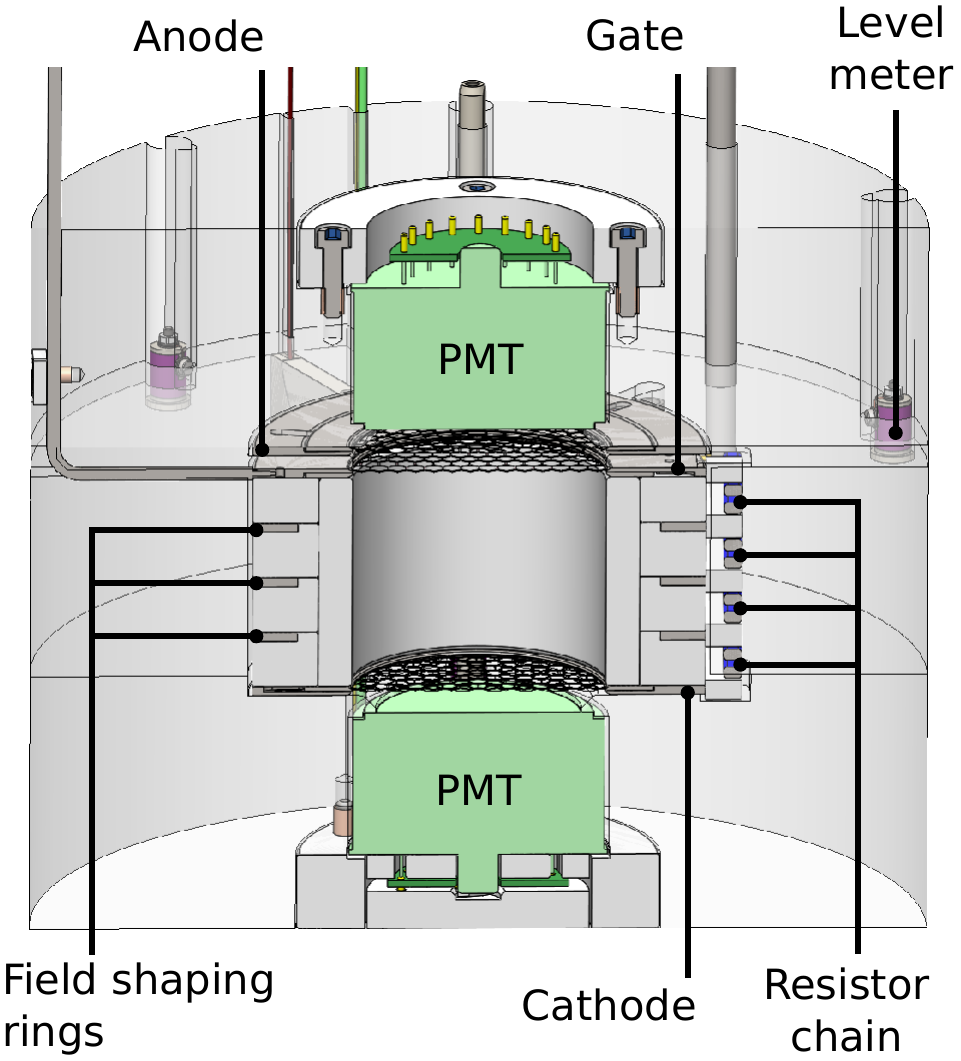}
		\caption{Schematic figure of the HeXe TPC with its components.}
		\label{fig:hexe_tpc_scheme}
\end{figure}

The HeXe TPC has a cylindrical drift volume of 5\,cm height and 5.6\,cm diameter containing a total mass of 345\,g liquid xenon (see figure\,\ref{fig:hexe_tpc_scheme}). 

It is instrumented with 2 Hamamatsu R6041-406 2-inch metal channel dynode photomultiplier tubes (PMTs) specially selected for a high quantum efficiency (QE of 39.1\% and 38.8\% for top and bottom, respectively). The PMTs are typically operated between $(900-980)$\,V with a negatively biased high voltage (HV) being provided by an iseg power supply. A custom-made voltage divider 
is used to distribute the HV to the PMT cathode and dynodes. The single photoelectron (SPE) response of the PMTs can be measured \textit{in situ} using external LED boards, illuminating the two PMTs with photons via two optical fibers.	
Due to the large diameter of the cryostat hosting the TPC, a PTFE support structure has been designed in which active volume is embedded. In this way, the amount of LXe required to fill the active volume to the desired height is minimized. The active volume is
surrounded by an easily exchangeable PTFE cylinder, which can be used to study the impact of PTFE chemical cleaning on the purity of liquid xenon, as has been reported in\,\cite{Bruenner:2020arp}.

\vskip 0.02cm

The electric fields in the TPC are generated using three hexagonal pitched grids as described in section\,\ref{sec:efield}. The drift field is homogenized using a field cage assembly consisting of three field shaping rings. Cathode and gate electrodes are connected to them by a resistor chain using four $1\,\mathrm{G\Omega}$ high voltage resistors with a tolerance of 1\%.
A custom made CF40 flange was designed as HV feedthrough. It is equipped with three KINGS SHV 10\,kV welded HV connectors insulated with an additional PTFE piece. 
\vskip 0.02cm

In order to measure the large signals originating from alpha energy depositions, the emitted light needs to be attenuated before illuminating the PMTs to avoid both DAQ and PMT-base saturation. The attenuation is achieved by using thin PTFE sheets of 700\,$\upmu$m and 550\,$\upmu$m thickness in front of the top and bottom PMT, respectively.
The transparency of these PTFE sheets to liquid xenon scintillation light was determined separately and is reported in~\cite{Cichon:2020ytl}.


\section{Electric field configuration}
\label{sec:efield}

The electric fields in the TPC are maintained using three electrode meshes.
The drift field which guides the electrons upwards, out of the interaction site, is defined by the cathode and gate electrodes.
While the drift field has been varied in the measurements shown here, the field applied to extract electrons from the liquid to the gas phase and amplify the charge signal\,\cite{Lansiart:1976} is kept at an approximate value of $10$\,kV/cm in the gas phase. 
High voltage is applied to the electrodes using two iseg modules, where one of them provides a high-precision current read out allowing for monitoring of the cathode-to-gate current.
\vskip 0.02cm

The electrode meshes have a hexagonal pattern and were etched out of 100\,$\mathrm{\upmu m}$ thick stainless steel sheets by Great Lakes Engineering.
Each of the electrodes is housed inside a 2\,mm thick circular stainless steel frame having an outer diameter of 10\,cm and an inner diameter of 5.6\,cm. The gate and cathode electrodes are rotated horizontally by $\pm 45^\circ$ with respect to the anode electrode, due to construction reasons. Like the electrode frames, the three field shaping rings are made from 2\,mm thick stainless steel with an outer diameter of 10\,cm and an inner diameter of 8\,cm and are placed equidistantly along the drift length. Voltage is applied to each electrode separately using three PTFE insulated cables guided along the outer region of the PTFE support structure.


\subsection{Electric field simulation}
\label{subs:E_sim}

The design of the HeXe TPC has been optimized for a high electric-field homogeneity based on a detailed detector simulation.
Using the commercial finite element software COMSOL Multiphysics\textsuperscript{\tiny\textregistered}\,\cite{comsol:1998}, a model containing all necessary electrical components was created. The number of field shaping rings or the design of the electrodes (section\,\ref{subs:electrodes}) were optimized based on the outcome of this simulation.
\vskip 0.02cm

\begin{figure}[h]
	\centering
	\includegraphics[width=0.48\textwidth]{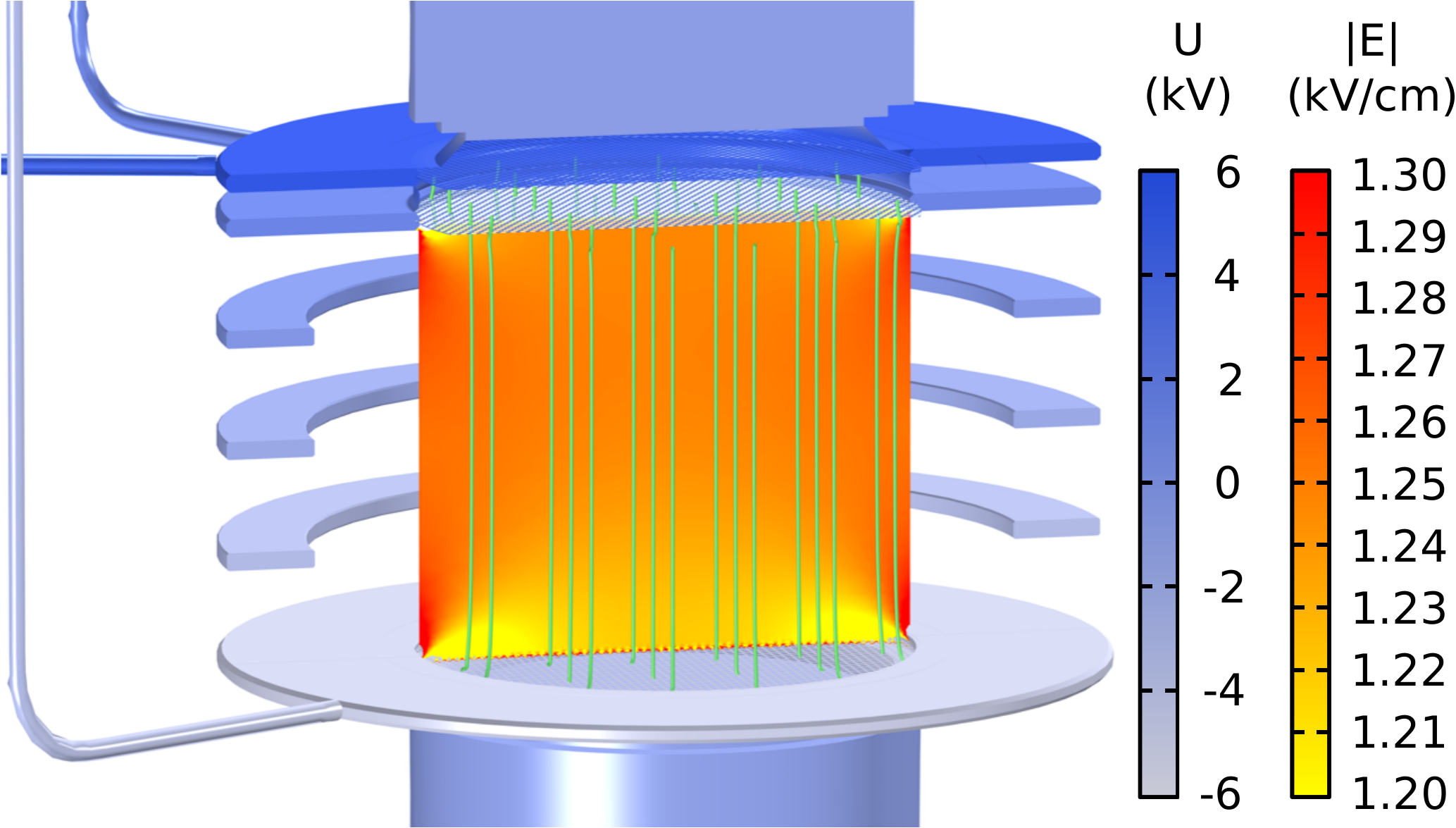}
	\caption{Illustration of the COMSOL model used for electric field simulation. Shown is the outcome for an average field of 1.25\,kV/cm. While the blue color scale represents the potential at the metallic surfaces, the orange scale shows the field intensity across a plane through the central region. Several selected electric field lines are illustrated in green.}
	\label{fig:e_field_3d_model}
\end{figure}

For a correct representation of the hexagonal pitched electrodes as well as their relative rotation, a full three-dimensional geometry is chosen instead of the more common 2D approximation. 
Figure\,\ref{fig:e_field_3d_model} shows the 3D geometry as implemented in COMSOL. The blue color scale displays the value of the potential at each metallic surface. The colored plane represents the intensity of the field in the target region. The complete field map, also outside the TPC, is shown in figure\,\ref{fig:field_map_pitches}.

In order to calculate the field in the TPC for any given combination of electrode and PMT voltages, the superposition principle can be used.
First, the field distribution stemming from each individual electrical component is simulated with a potential value of one volt with all other components set to ground potential.
These results are then exported from COMSOL on a grid of 100\,x\,100\,x\,100 points, evenly spaced in $(z,r^2,\varphi)$ throughout the active volume.
Afterwards, the overall field can be computed efficiently by summation of the individual fields, scaled to the desired voltage of each respective component.
This approximation has been validated against the result of a direct field computation, and it is accurate on the \textit{per mille} level for the majority of the active volume.
The superposition method avoids the computationally expensive re-computation of the full solution for each voltage combination, allowing for an iterative optimization of the field homogeneity with respect to the voltages applied to the electrodes.
Residual field inhomogeneities are quantified using the central 68\% inter-percentile range of the field magnitude from all points of the grid.

\subsection{Electrode design}
\label{subs:electrodes}

The electrodes should provide a high homogeneity of the electric field while having the maximum possible optical transparency in order to maximize the collection of photons.
The influence of the electrode design onto the homogeneity of the electric field is evaluated using the detector field simulation.
The electrodes are modeled, as the rest of the detector, in three dimensions assuming a rectangular wire cross-section. The influence on the optical transparency is estimated using the geometrical coverage.
\vskip 0.02cm

For the same geometrical coverage, electrodes made from parallel wires allow for a more narrow pitch, which in turn leads to a slightly higher field uniformity when compared to an hexagonal shaped electrode having the same geometrical coverage.
However, given its higher mechanical stability, the hexagonal design has been favored for the HeXe setup.
For these electrodes, several combinations of the short diagonal length of the hexagonal cells (pitch, p) and wire thickness (d) were compared within ranges of $\mathrm{p} = \left[1\,\mathrm{mm},\,..., 10\,\mathrm{mm}\right]$ and $\mathrm{d} = \left[0.1\,\mathrm{m m},\,..., 1\,\mathrm{m m}\right]$ respectively.
\vskip 0.02cm

\begin{figure}[h]
	\centering
	\includegraphics[width=0.5\textwidth]{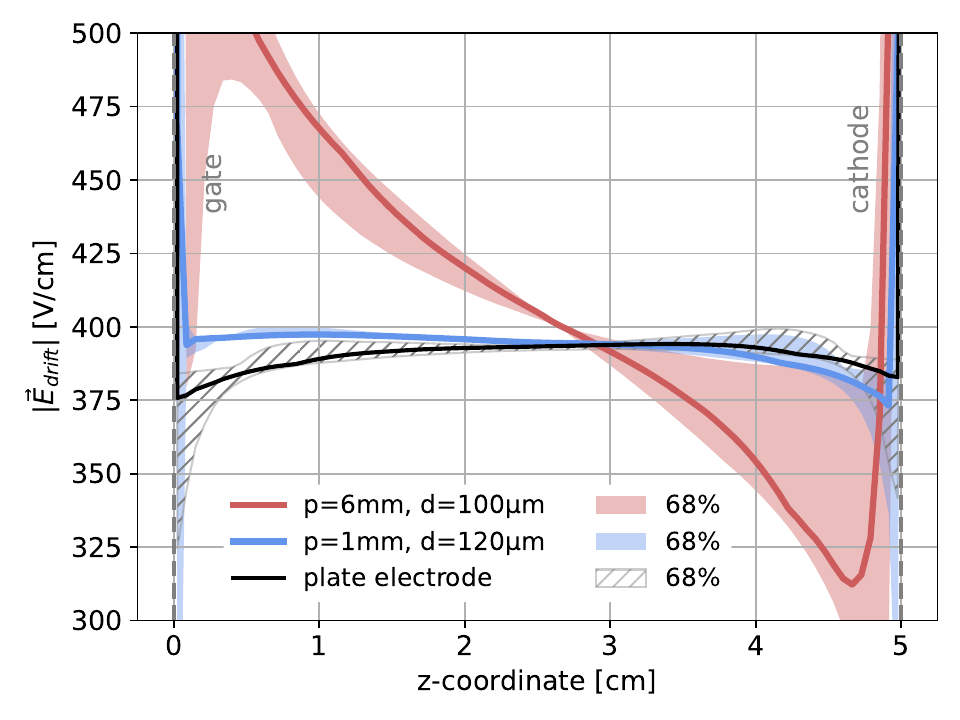}
	\caption{Electric field as function of depth ($z$-coordinate) in the TPC for different simulated values of wire thickness and pitch, in comparison to ideal plate electrodes (black line).}
	\label{fig:Efield_pitches}
\end{figure}

Figure\,\ref{fig:Efield_pitches} compares the simulated magnitude of the electric field as a function of the $z$-coordinate for the two types of electrodes that have also been tested in the HeXe TPC.
The lines indicate the median of the absolute value of the electric field, whereas the shaded regions show the range of 68\% field deviation within each slice along the height of the active volume.
The field that would be exerted by electrodes made from solid plates is shown in black. Since in this case there would be no field leakage through the electrodes, it can be regarded as the limiting case of best achievable field homogeneity with the setup.
The residual inhomogeneities of about 2.1\% are then due to stray fields which are not compensated by the field-shaping rings. 
Compared to this ideal case, hexagonal electrodes with a pitch of 1\,mm (blue line) show only a small amount of additional field inhomogeneities of 2.2\% while each electrode allows for a light transmission of 79.7\% at the same time.
For electrodes having a pitch of 6\,mm (red line), this transparency increases to 96.7\%, however the field also becomes significantly less uniform (29.5\%).
This holds true for homogeneity along the z-direction as well as in the x-y direction.
The transparencies of the hexagonal electrodes are estimated as the ratio between the uncovered surface and the total surface of the grids.
Note that by this approximation any dependence on the direction of the incoming light is not considered.
\vskip 0.02cm

\begin{figure}[h]
	\centering
	\includegraphics[width=0.49\textwidth]{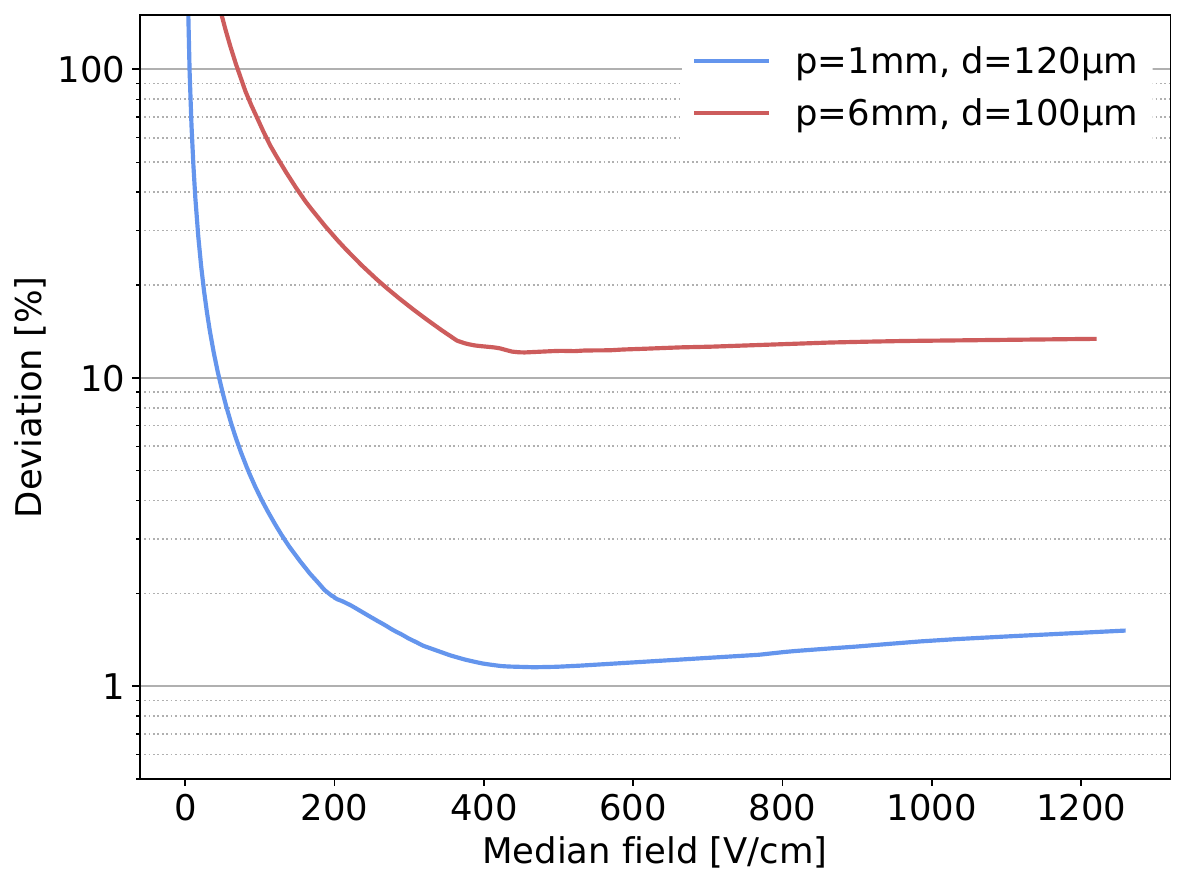}
	\caption{Dependence of the electric field uncertainty (68\% relative deviation)  as a function of the applied drift field for two electrode designs. As expected, the simulated field for smaller pitched electrodes (blue line) shows a significantly smaller variation. Regions close to the electrodes are removed by a similar fiducial volume selection as used for the analysis.}
	\label{fig:Efield_scan}
\end{figure}

Figure\,\ref{fig:Efield_scan} shows the field deviation from the median field value as a function of the applied field strength.
It can be seen that the relative uncertainty quickly increases for very small drift fields. This is due to the unavoidable field leakage from the extraction region in which a strong electric field needs to be maintained.
At a median drift field of 460\,V/cm the relative deviation reaches a minimum of 1.2\%. At higher applied fields, the deviation slowly increases. We assume this effect comes from the field leakage from the region below the cathode. The deviation reaches a value of 1.5\% for a drift field of 1.26\,kV/cm.
While the drift field is being changed during the measurement, the extraction field needs to be kept constant.
It is estimated to vary by only 0.5\% over the full range of applied drift fields which can be assumed to have a negligible influence on the detected S2 size.
\vskip 0.02cm

\begin{figure*}[h]
	\includegraphics[width=0.97\textwidth]{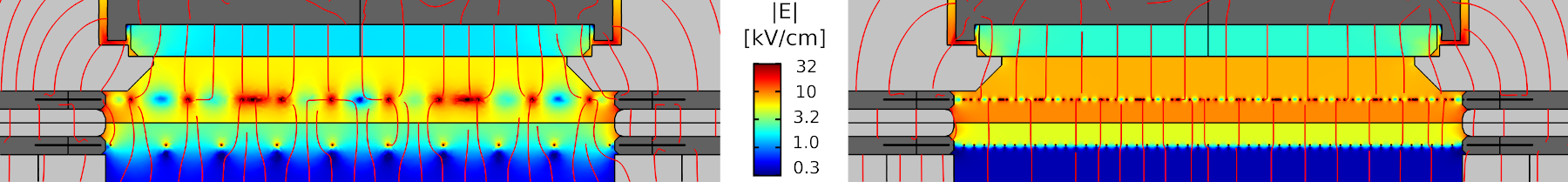}\\[1em]
    \includegraphics[width=0.97\textwidth]{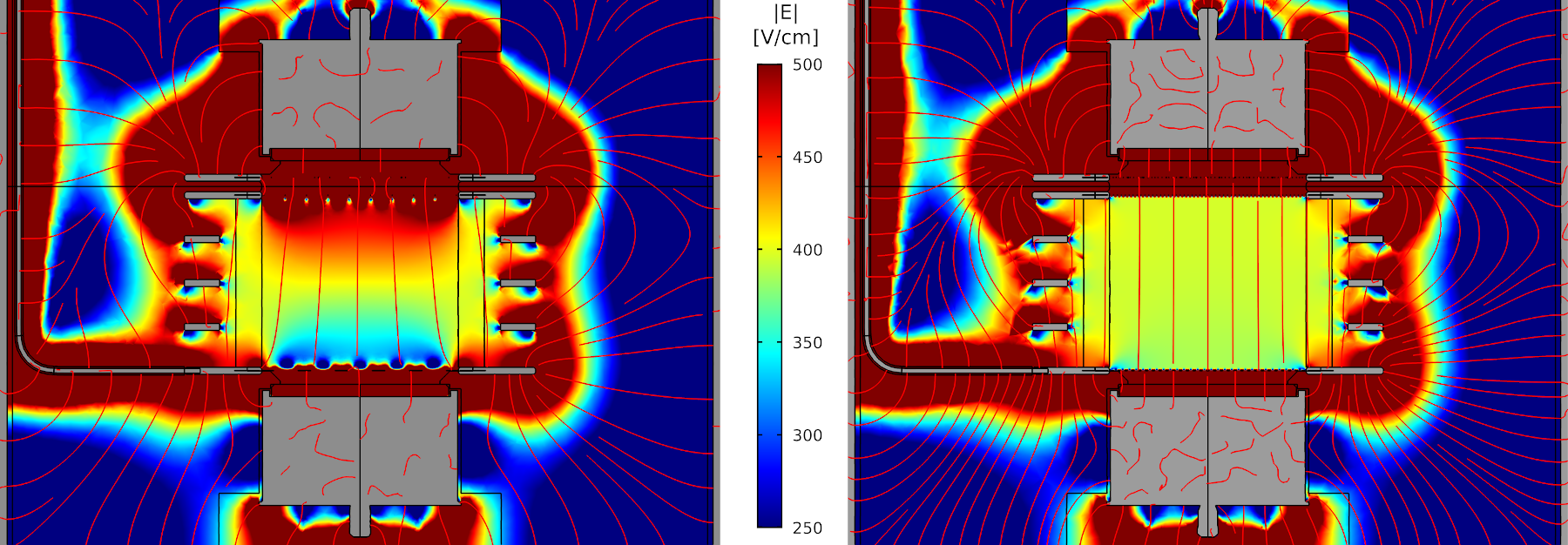}\\[1em]
    \includegraphics[width=0.97\textwidth]{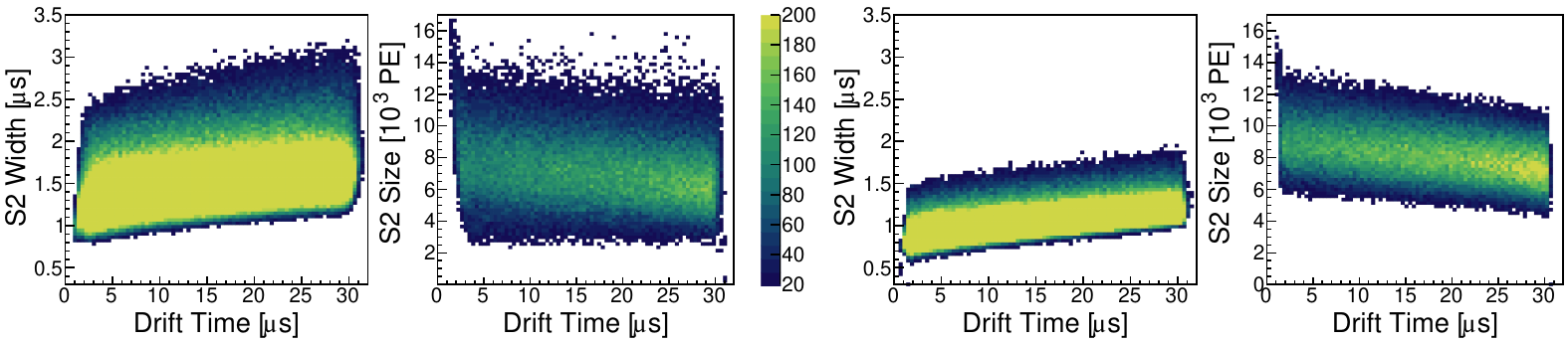}
	\caption{Simulation results and data from ${}^\mathrm{83m}\mathrm{Kr}$ decays for hexagonal grids of 6\,mm pitch (left) and 1\,mm pitch (right). The voltages of the individual components are -900\,V, +3750\,V, -5\,V, -1950\,V and -950\,V for the top PMT, anode, gate, cathode and bottom PMT respectively, yielding a median electric drift-field of about 400\,V/cm. \textbf{Top:} Zoom into the extraction region of the HeXe TPC. Shown is the magnitude of the electric field on a logarithmic scale. Steel and PTFE components are colored in dark and light gray, respectively. \textbf{Center:} Magnitude of the electric field inside the drift volume as well as within the PTFE components of the TPC. Steel components are illustrated in gray. \textbf{Bottom:} Comparison of the S2 width and S2 area resolution in ${}^\mathrm{83m}\mathrm{Kr}$ data acquired with grids of 6\,mm and 1\,mm pitch.} \label{fig:field_map_pitches}
\end{figure*}

The two grid options compared in figures\,\ref{fig:Efield_pitches} and\,\ref{fig:Efield_scan} were also installed in our TPC.
In a first configuration the meshes had a wire thickness of $100\,\upmu $m and a pitch of 6\,mm and in the second configuration $120\,\upmu $m and 1\,mm, respectively.
Figure\,\ref{fig:field_map_pitches} compares the outcome of the simulation for the two configurations (6\,mm pitch on the left and 1\,mm on the right) and shows how it impacts the data quality from ${}^\mathrm{83m}\mathrm{Kr}$ decays in the TPC (see also section\,\ref{subs:data_analysis}) at a median field of approx. 400\,V/cm. 
While the middle panels show the field in and around the target, the upper panels are zoomed into the electron extraction region showing the two electrodes and the quartz PMT window.
The color scale represents the magnitude of the electric field. 
The lower panels show for each case data of the S2 pulse width (left) and the S2 pulse size (right) as a function of the drift length in the TPC. In the liquid-xenon target region, the second configuration (right side in figure\,\ref{fig:field_map_pitches}) gives a significantly more homogeneous electric field. The deviation from the median field is $(1-2)\%$ for the 1\,mm pitch and $\sim 13$\% for the 6\,mm pitch (see figure\,\ref{fig:Efield_scan}).
\vskip 0.02cm

The increased field homogeneity in the case of the electrode configuration with the smaller pitch results in an enhanced quality of the data. The variation in the S2 pulse width is strongly reduced with a standard deviation changing from 0.40 to 0.23 in the drift region $(10-12)\,\upmu $s for the 6\,mm and 1\,mm pitch, respectively. Similarly, the S2 size distribution is narrower (see section\,\ref{sec:DAQ_processing_analysis} for more details on the data analysis).
Note that the S2-width increases with the drift distance due to diffusion and the S2-size decreases as a result of electron attachment to impurities in the liquid xenon. The correction for the latter effect is described in section\,\ref{sec:DAQ_processing_analysis}. For all measurements shown in this paper, we use grids with d\,=\,120\,$\upmu$m and p\,=\,1\,mm for all three electrodes.


\section{Auxiliary systems}
\label{sec:systems}


\subsection{Cryogenic, purification and calibration systems}
\label{subs:cryosystem}
	
The detector is operated inside a 20.1\,cm diameter vacuum insulated cryostat. Continuous cooling is provided by a Leybold COOLPAK 6000-1 helium-driven pulse tube refrigerator (PTR). A copper cold head attached to it is used to condense the xenon. The temperature of the cold head is adjusted to $T_{\mathrm{cold~head}} = -107.5\,\mathrm{^\circ C}$ using a PID controlled heater cartridge. This cryogenic system was already operated successfully in previous measurements\,\cite{Barrow:2016doe,Bruenner:2016ziq}.
	\vskip 0.03cm
	
\begin{figure}
	\centering
	\includegraphics[width=0.48\textwidth]{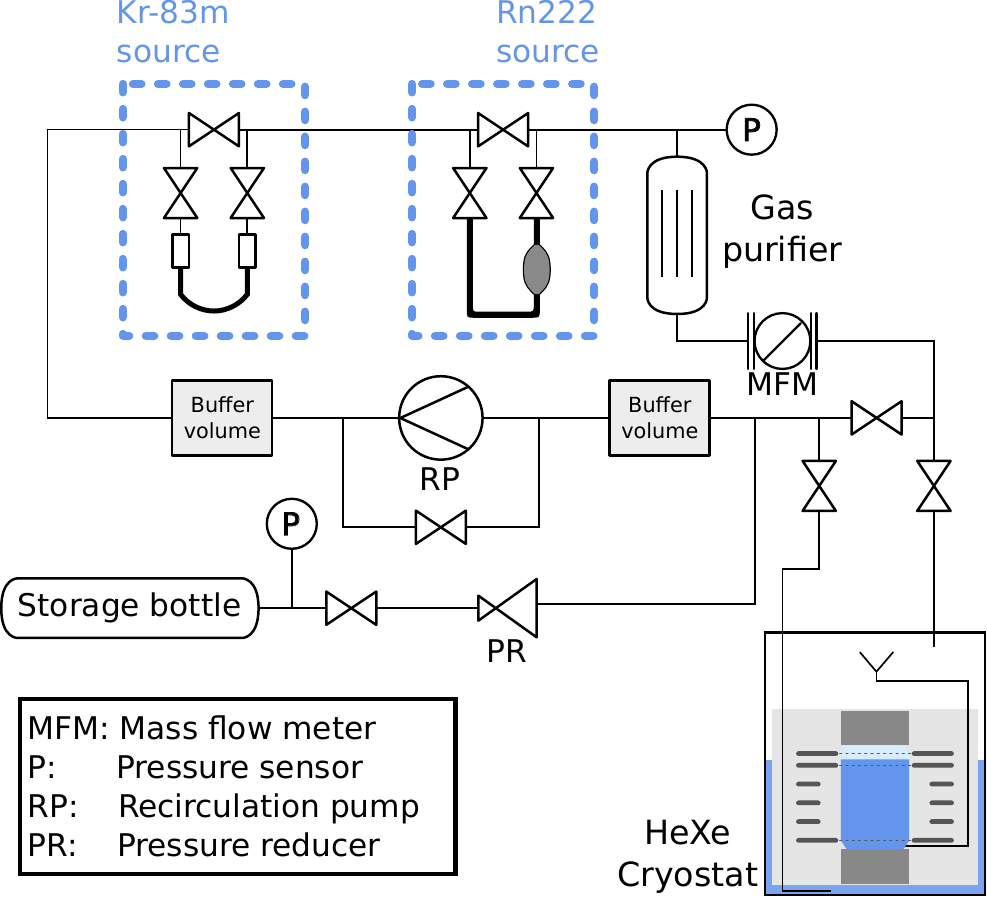}
	\caption{Simplified scheme of the HeXe gas system used for xenon purification and introduction of gaseous calibration sources.}
	\label{fig:hexe_gas_system}
\end{figure}

A gas handling system is employed to store the xenon, to introduce calibration sources into the detector, and to remove electronegative impurities from the xenon. The last is necessary to maintain sufficiently large light and charge yields as impurities like oxygen and water decrease the signal yields.  The purification is obtained by recirculating the xenon gas with a KNF double-diaphragm pump through a hot SAES gas purifier \cite{SAES} which continuously removes  electronegative impurities from the xenon with a very high efficiency.
A schematic representation of the gas system is shown in figure\,\ref{fig:hexe_gas_system}. 
In the purification loop, the liquid xenon is extracted from a reservoir below the bottom PMT. Simultaneously, clean xenon is introduced to the gas phase where it liquefies at the cold head. The liquid drops are subsequently guided back to the liquid phase through a funnel and enter the detector through a channel located above the bottom PMT window. The gas recirculation flow is continuously monitored by a MKS mass flow meter and controlled to a value about 2.75\,SLPM using a valve bypassing the recirculation pump.
\vskip 0.03cm

Opening the cryostat and performing modifications on the TPC that are relevant for the measurements (e.g. installation of light attenuators) are carried out while minimizing the exposure to air\,\cite{Bruenner:2020arp}.  These operations are performed in a permanently installed glove bag under a GN$_{2}$ atmosphere at a slight over-pressure.
This procedure reduces the contamination from gas components present in ambient air that could diffuse into e.g. the PTFE of the TPC and reduce the xenon purity.
	\vskip 0.03cm
	
Two calibration sources are employed for the measurements of sections\,\ref{sec:response} and~\ref{sec:drift}: a ${}^\mathrm{83m}\mathrm{Kr}$ electron-emitting source and a $^{222}$Rn alpha source. Both are introduced into the TPC through the gas system as shown in figure\,\ref{fig:hexe_gas_system}. ${}^\mathrm{83m}\mathrm{Kr}$ is emanated from ${}^\mathrm{83}\mathrm{Rb}$ loaded zeolite beads\,\cite{lebeda_rb83} and produces internal conversion electrons with energies of 32.1\,keV and 9.4\,keV. The rate of krypton inside the target varies between a few kHz to $\sim100$\,Hz due to the short half life of the $^{83}$Rb source producing ${}^\mathrm{83m}\mathrm{Kr}$.
The alpha source originates from a liquid $^{226}$Ra standard. The emanated $^{222}$Rn is extracted from the source using a flow of helium and stored in an activated carbon trap. The trap is subsequently attached to the gas system and can be opened to the xenon circulation flow via the opening of a bypass.
Due to the $^{222}$Rn half-life of almost 3.8 days, it takes at least 3 weeks for the detector to reach its background level after the introduction of the source. Typically, a rate about $100$\,Hz is introduced into the liquid xenon target.

	
\subsection{Slow control system and detector stability}
\label{subs:sc_daq}
	
The gas pressure inside the cryostat, in the gas system and in the cryostat's insulation vacuum are measured using digital manometers and vacuum sensors. The temperature inside the TPC is recorded with four Pt100 thermometers which provide as well a rough estimate of the liquid xenon height during filling. A more precise measurement of the liquid level is made by four custom-made level meters which are distributed across the circumference of the support structure. Consisting of a concentric assembly of an inner $1$\,mm thick metal rod surrounded by an outer cylinder of $4$\,mm diameter, they allow to determine the LXe height by relative capacitance increase. Three of the level meters have a length of 1\,cm and are sensitive to the liquid xenon levels between the gate and the anode region. The fourth level meter, having a length of 10\,cm can be used to monitor the complete height of the drift volume.
 \vskip 0.02cm

All sensors are read out by a LabVIEW\,\cite{Labview} program, with the values stored into a periodically backed-up PostgreSQL\,\cite{Postgresql} database for later access. They are running on the same machine to ensure database input in case of network outages.  For each sensor, two separate sets of thresholds can be specified (warning and alarm). If the readings pass them, emails or SMS messages are being sent to HeXe operators to alert them. Both the program and the database are, in turn, monitored using Nagios\,\cite{Nagios} which is set up on a separate server, in order to also send alerts in case at least one of them is not responsive or accessible.
The modules providing HV for the PMTs and grids are controlled by a tool written in Python. It continuously reads out voltage and current values, writing them into the above-mentioned database. Furthermore, it is capable of notifying operators of HV trips via email/SMS and attempts to automatically ramp up tripped channels if desired. In addition, it monitors the cryostat pressure via the database. If the pressure drops below a customizable threshold, all HV channels are ramped down to avoid damage due to discharges. Schedules to ramp up or down HV can also be set making automatic voltage scans possible.
\vskip 0.02cm

\begin{figure}
	\centering
	\includegraphics[width=0.48\textwidth]{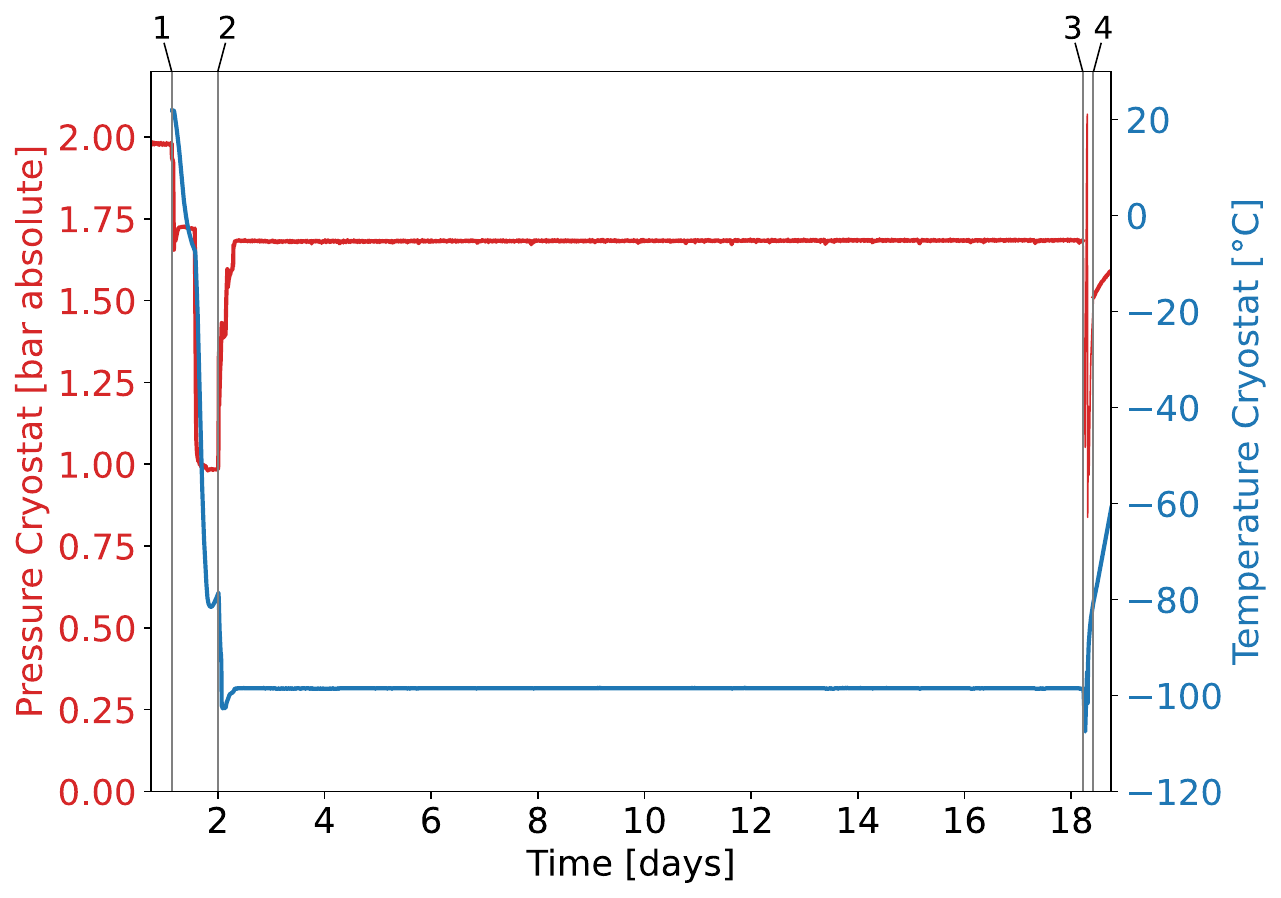}
	\caption{Evolution of temperature (in blue) and pressure (in red) during the $\mathrm{^{222}{Rn}}$ measurement. After filling the cryostat, the cool-down phase is started (1). Additional GXe is filled and liquefied (2). After reaching the required LXe level, the xenon recirculation is started and the measurement begins. The measurement ends with the recuperation of the LXe (3). About 1\,bar of GXe is left in the cryostat and recirculated, to speed up the warming-up time (4).}
	\label{fig:TP_evol}
\end{figure}

Figure\,\ref{fig:TP_evol} shows the pressure and temperature evolution inside the cryostat during the $\mathrm{^{222}{Rn}}$ measurement described in sections~\ref{sec:response} and \ref{sec:drift}.
A typical measurement starts by filling the cryostat with about 2\,bar (absolute pressure) of GXe.  Subsequently, the gas is cooled down (1) in two steps to approx. $-107.5 ^{\circ}$C (at the cold head).
 Next, GXe is filled and rapidly liquefied in the pre-cooled system (2). To reach the required xenon level inside the TPC a total mass between 2.5 and 3\,kg is necessary. In this state, the gas recirculation starts and the measurement period begins. During the stable operation period, between 2.5 and 17.5 days, the average pressure and temperatures are $(1.68 \pm 0.02)$\,bar and $ (-98.5 \pm 0.02)$\,$^{\circ}$C, respectively. The run is terminated when the xenon gets cryo-pumped into the storage bottle\,(3). This causes a pressure decrease down to an adjusted gas amount of 1\,bar inside the cryostat. Xenon gas is recirculated in the gas system to speed up the warming up process (4).

	
\section{DAQ, data processing and data analysis}
\label{sec:DAQ_processing_analysis}
	
\subsection{Data acquisition system}
\begin{figure}[h]
	\centering
	\includegraphics[width=0.47\textwidth]{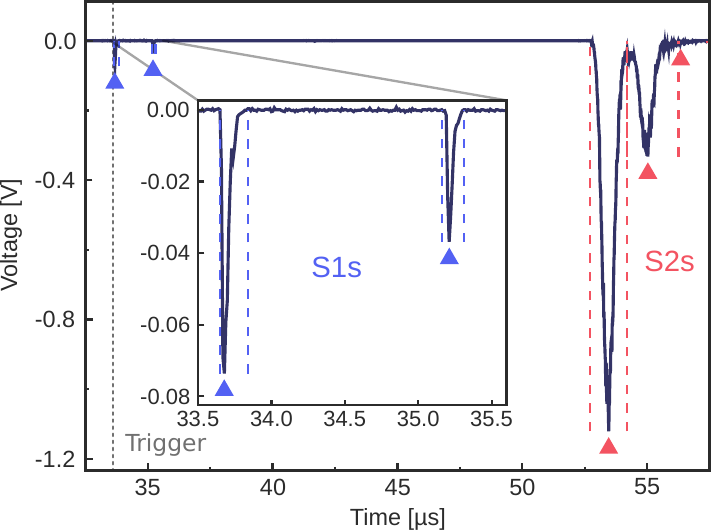}
	\caption{$^{83\textrm{m}}$Kr candidate event. The signals and the intervals as identified by the data processor are shown. The two main S2 signals are followed by a small single-electron S2 signal which is caused by delayed electron extraction or ionization of impurities inside the TPC.}
	\label{fig:kr_example_event}
\end{figure}

The data acquisition system (DAQ) takes care of digitizing PMT waveforms. The signals are taken directly from the HeXe signal feedthroughs to a custom-made voltage amplifier with an amplification factor of $\sim$2 per channel. They are then fed into discriminators with a threshold at about --10\,mV. The coincidence of two discriminators above their threshold gives a trigger. 
Triggered data is digitized with a CAEN V1724 module which has a sampling rate of 100\,MS/s, 14\,bit resolution and a dynamic range of 2.25\,V. A custom-made software is employed to configure the module, to control data acquisition and to store the recorded waveforms on disk.
Figure\,\ref{fig:kr_example_event} shows an example waveform of a ${}^\mathrm{83m}\mathrm{Kr}$ event in HeXe. The inset shows a zoom into the S1 signal region with the signals corresponding to the two characteristic decay lines.
On the right side, the signals from the two corresponding S2s can be seen. Due to the large width of the S2s, most of the two S2s from the ${}^\mathrm{83m}\mathrm{Kr}$ decay do however overlap. The approximate trigger position is indicated by the gray line.
\vskip 0.02cm

\subsection{Data Processing}
To process the acquired data, we use the software described in~\cite{Cichon2020}. By calculating the average of the first 25 available samples, the waveform baseline is estimated.
The baseline-subtracted waveforms from both PMTs are added to a summed waveform which is used for analysis. The standard deviation of the first 25 samples in the summed waveform, $\sigma$, is also computed. Next, peak candidate regions are found by looking for excursions which surpass $5\sigma$.
Afterwards, a threshold-based peak finding algorithm, which is detailed in~\cite{Cichon2020}, is applied. For each peak found, its type \textemdash\,S1 or S2 \textemdash\,needs to be determined. Various pulse parameters, such as rise-time or pulse width, are employed for this purpose. For S1 pulses, we require that the rise-time, defined as the time between the 0.1 and 0.5 area quantiles, is below 200\,ns and the pulse width (time duration between 0.1 and 0.9 area quantiles) to be smaller than 500\,ns. Peaks which do not satisfy these criteria are classified as S2 pulses. The 0.5 area quantile is taken as the peak's time relative to the beginning of the event in which it is found.
The efficiency of this criterion for S1 pulses from alpha interactions is evaluated to be $>99.9$\% using the pulse-shape parameters measured in\,\cite{Cichon:2022}.
\vskip 0.02cm

\begin{figure}[h]
	\centering
	\includegraphics[width=0.48\textwidth]{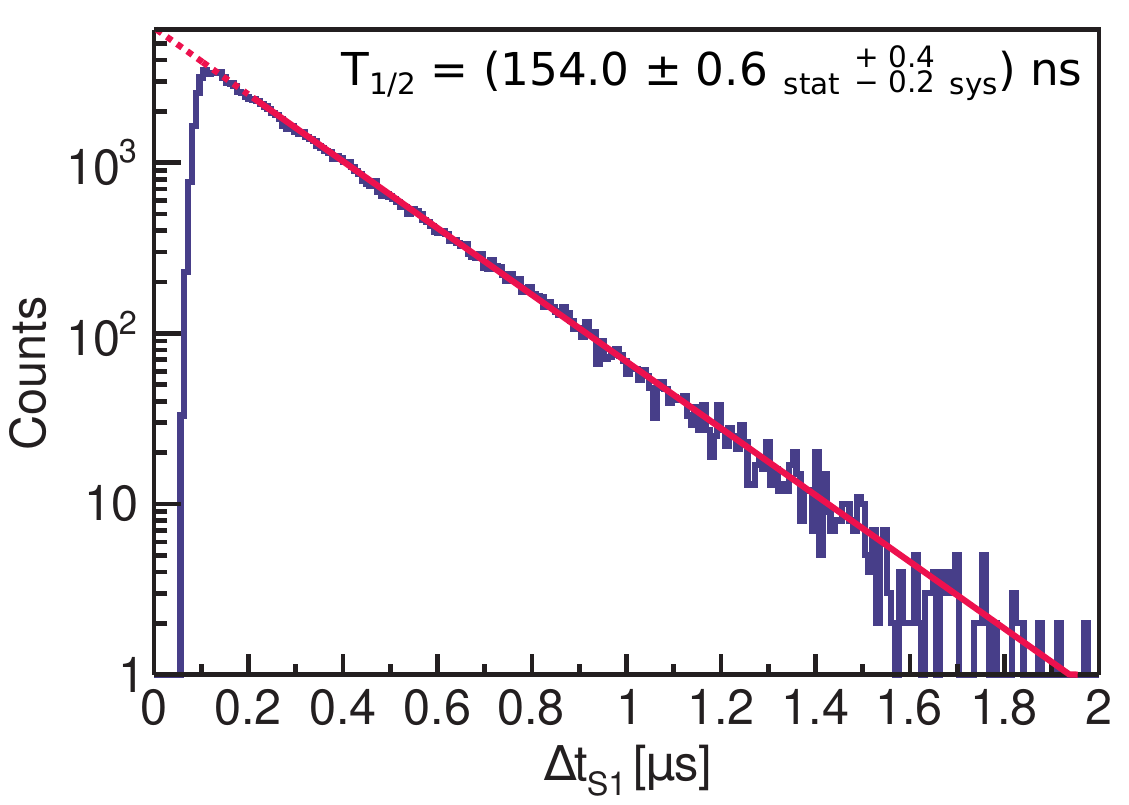}
	\caption{Time difference between the scintillation pulses (S1s) originating from the decays of $^{83\textrm{m}}$Kr at 32.1\,keV and 9.4\,keV.}
	\label{fig:Delta_t_S1s}
\end{figure}

Using the threshold-based peak splitting, an excellent separation between the two S1s is achieved. S1s as close as 50\,ns from each other can be resolved as shown in figure\,\ref{fig:Delta_t_S1s}. The figure shows the number of $^{83\textrm{m}}$Kr events with two separated peaks for different decay times. For times between the S1s smaller than about 125\,ns, the efficiency of separating both lines decreases. An exponential probability density function is fit to the distribution of time differences between both S1s in order to determine the half-life of the intermediate 9.4\,keV state. This is done by minimizing the unbinned negative log-likelihood in the range between 0.2\,µs and 2\,µs. The systematic error is estimated by varying the lower fit bound by $\pm 75\,\mathrm{ns}$. From the fit, a value of 
$T_{1/2} = \left(154.0 \pm 0.6\,_{\mathrm{stat}}\,^{+0.4}_{-0.2}\,_{\mathrm{sys}}\right)\, \mathrm{ns}$
is extracted which is compatible with previous publications~\cite{Ahmad:1995zz}.
The small oscillations on top of the exponential distribution are caused by high-frequency noise in the signal of the top PMT.

\begin{figure}[h]
	\centering
     \includegraphics[width=0.45\textwidth]{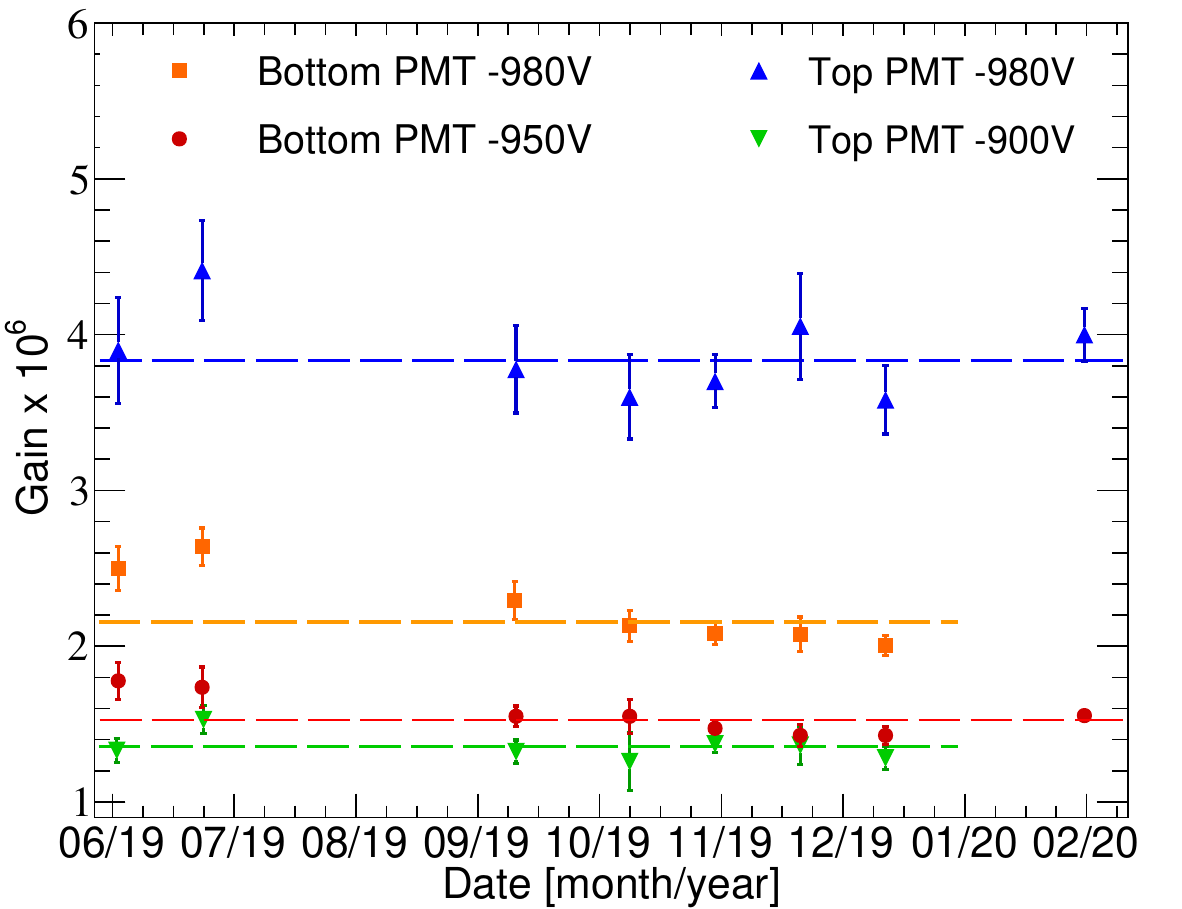}
	\caption{Stability of the top and bottom PMT gains for two different HV configurations. The dashed lines are constant fits to the data.}
	\label{fig:PMT_stabil}
\end{figure}

The PMT gain is defined as the average charge generated per photoelectron (PE)
in units of the elementary charge e$^{-}$. In each run, the PMT gains are calibrated using photons from a 350\,nm LED located on trigger boards outside the LXe chamber. Two optical fibers feed the light into the active volume of the TPC, 20\,mm below the gate. The same model-independent method as described in~\cite{Saldanha:2016mkn} is employed to regularly determine the gain of the PMTs. Figure\,\ref{fig:PMT_stabil} shows the results of the gain calibration for the top and bottom PMT, 
each at two different HV configurations for the period between June 2019 and March 2020, with the dashed lines showing the fit of the data with a constant. 
To process the data, gain values measured right before the respective measurement are taken for the conversion of the peak area into units of PE.
\vskip 0.03cm

\subsection{Data analysis}\label{subs:data_analysis}

Alpha events from the $^{222}$Rn source are selected if precisely one S1 signal, followed by an S2 signal are found in the waveform. 
For $^{83\textrm{m}}$Kr events, we require that an S1 signal is followed by a smaller S1 and that at least one S2 signal occurs in time after the larger S1 signal.
Additionally, the S2 width needs to fit the expected dependence with drift time, and the fraction of the S1 signals seen in the top PMT needs to correspond to the region of physical events.

The distribution of S2 photons among the two PMTs is used to select high quality events as well. Due to an inclined Teflon piece at high radii on the top of our TPC (see top panel of figure\,\ref{fig:field_map_pitches}), the S2 area fraction seen by the top PMT has a characteristic double peak structure that can be used to effectively select an inner radius for analysis. 
A fiducial volume is defined by cutting in the $z$-coordinate events such that $10\,\textrm{mm}<z<35\,\textrm{mm}$. Especially for the two lowest fields investigated, the data quality improves with time due to an unknown reason. The statistics at the alpha line grows and the cathode position (see figure\,\ref{fig:GateCathode}) becomes sharper. For this reason, we take only the second half of the acquired data for the analysis of all fields below 60\,V/cm. Variations on this selection are performed to quantify the related systematic error.

\vskip 0.01cm

\begin{figure}[h]
	\centering
		\includegraphics[width=0.49\textwidth]{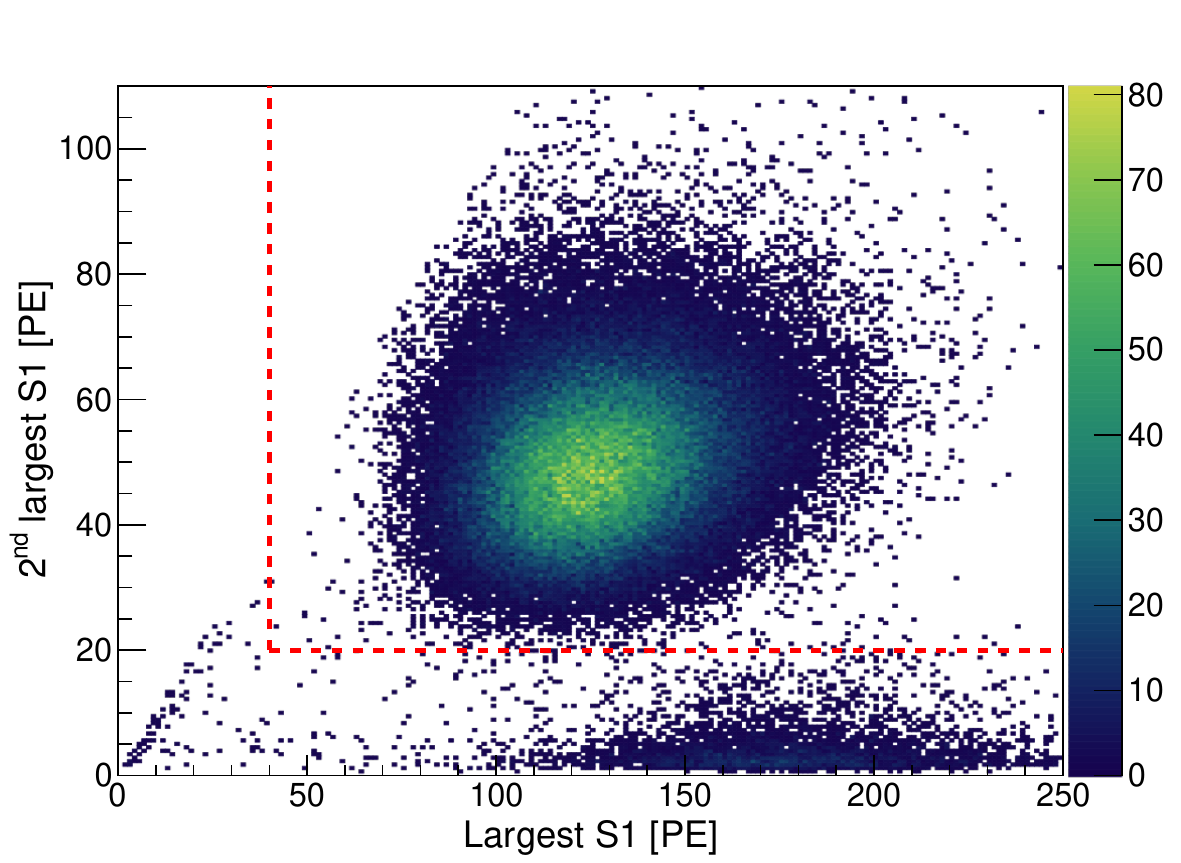}
	\caption{Population of $^{83\textrm{m}}$Kr events and the selection (red lines) using the time coincidence of its 32.1\,keV and 9.4\,keV decay lines.}
	\label{fig:Kr_Selec}
\end{figure}

\begin{figure}[h]
	\centering
    \includegraphics[width=0.49\textwidth]{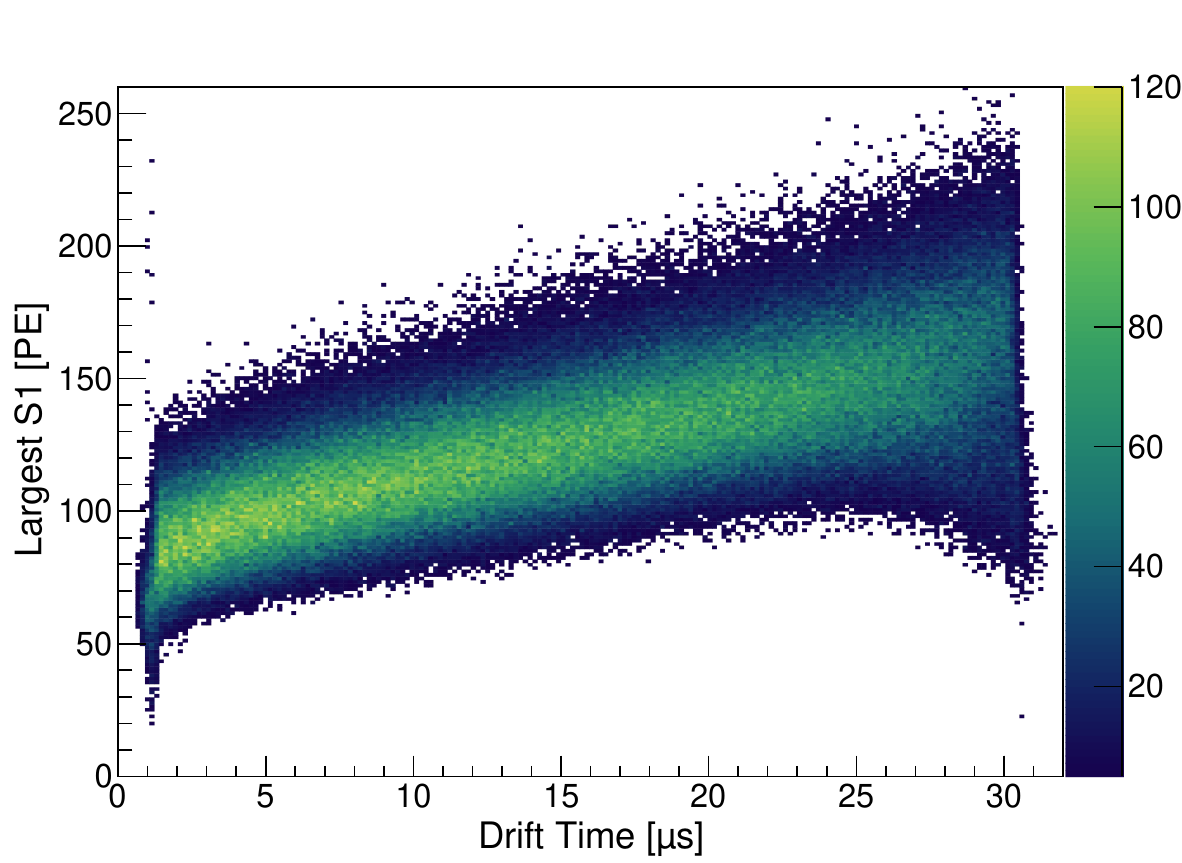}
	\caption{Dependence of the 32.1\,keV S1 from $^{83\textrm{m}}$Kr on the drift time due to the geometric variation of the light collection efficiency.}
	\label{fig:z-correctionS1}
\end{figure}

The size of the recorded S1 and S2 signals depends on the event location in the detector due to position-dependent effects, such as light collection efficiency (LCE) and electron attachment to impurities in the liquid xenon\,(see\,\cite{Aprile:2012vw} for details).
The signal sizes are corrected for the $z$-dependence of these effects. Two different corrections are necessary as the attenuators in the $^{222}$Rn measurements modify the light collection of each PMT differently than in $^{83\textrm{m}}$Kr.
The selection of $^{83\textrm{m}}$Kr events for a drift field of 394\,kV/cm is shown as an example in figure\,\ref{fig:Kr_Selec} (region above the red lines).
Figure\,\ref{fig:z-correctionS1} shows the dependence of the uncorrected S1 light yield on the drift time due to light collection effects. 
\vskip 0.03cm

The drift time dependence of the charge signal, where the signal loss is due to attachment of electrons to electronegative impurities in the liquid, can be seen in the lower right panel of figure\,\ref{fig:field_map_pitches}. It follows an exponential decay law, where the time-constant is usually referred to as the electron lifetime. To evaluate the $z$-dependence of the S2 signal size, the median value of the event distribution is determined for different slices along the $z$-coordinate. These values are then fit with an exponential function which is used to correct the data to the value at the top of the TPC where no losses are expected. For the datasets employed for the charge and light yield measurements, an average electron lifetime is used for the correction of each field. This is enough as each field is measured for just 4 hours.
Note that while the S1 $z$-correction can be determined once for the complete run, the S2 $z$-dependence is time dependent due to the continuous cleaning of the liquid xenon which results in a constantly improving S2 collection.
\vskip 0.03cm
 
\begin{figure}[h]
    \centering
    \includegraphics[width=0.49\textwidth]{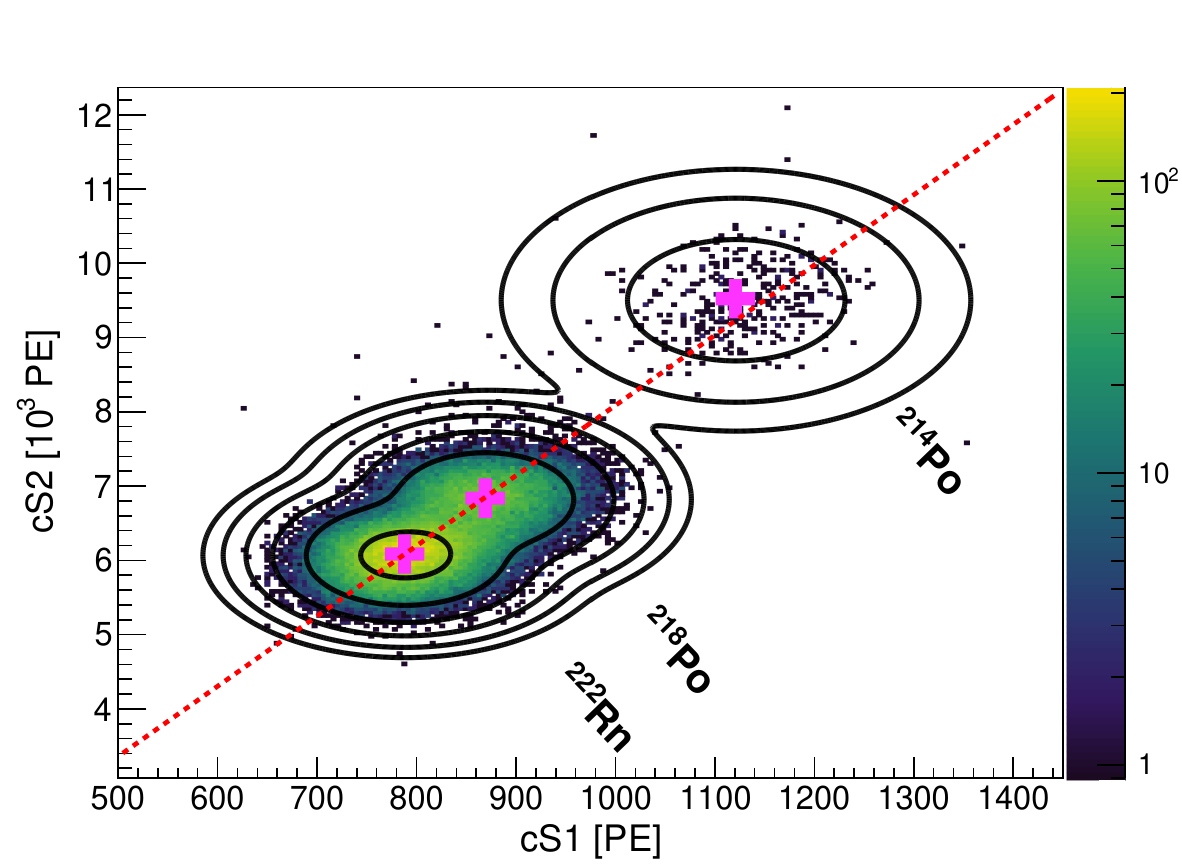}
    \caption{$^{222}$Rn data in the space of corrected S1 versus S2 at an electric drift-field of 1.24\,kV/cm.
    The estimated mean S1 and S2 signal sizes of the three isotopes are indicated by the pink markers.The contour lines of the fit (sum of three individual 2D-Gaussian functions) are drawn in black. The dashed red line shows the linear fit through the fitted mean positions.}
    \label{fig:rn222_sel}
\end{figure}

\begin{figure}[h]
    \centering
    \includegraphics[width=0.5\textwidth]{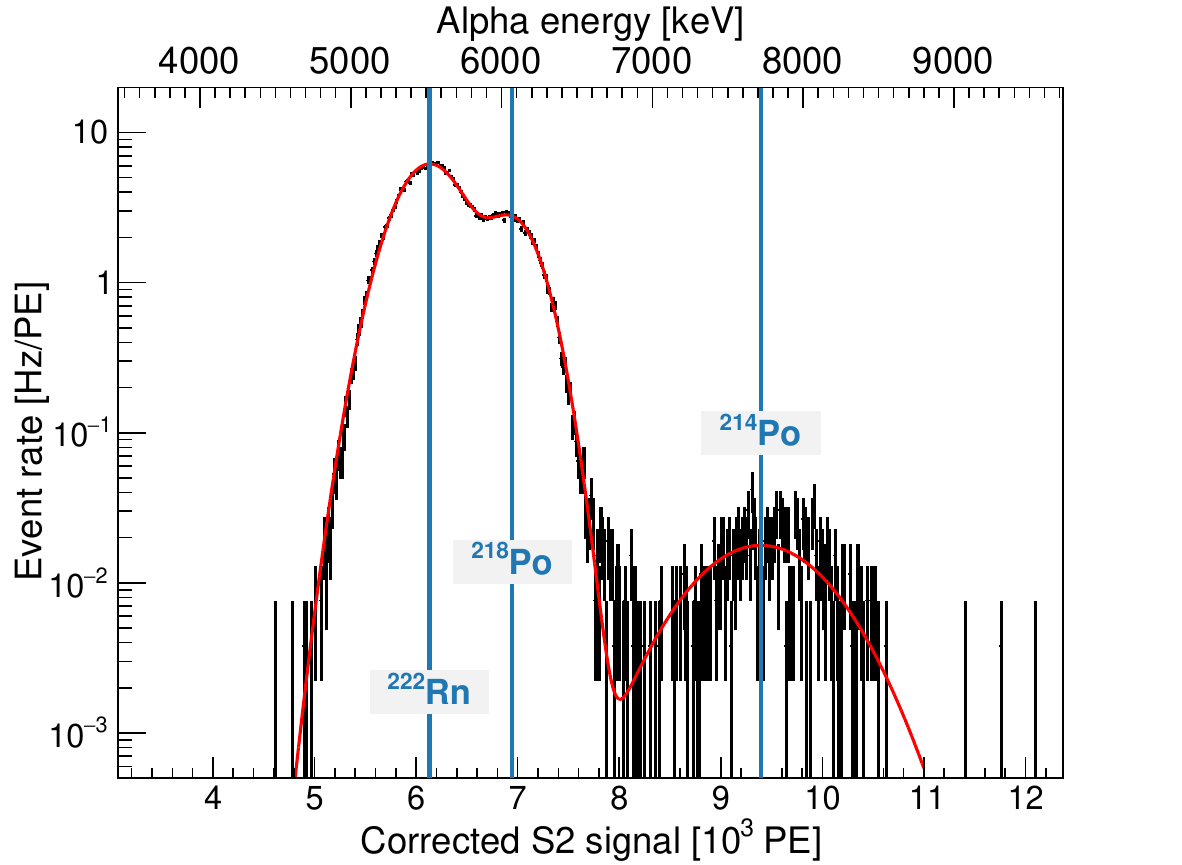}
    \caption{Corrected charge signal (S2) spectrum of the alpha decays from the $^{222}$Rn-calibration at an electric drift field of 1.24\,kV/cm.}
    \label{fig:rn222_cs2_spectrum}
\end{figure}

Alpha events are selected by looking at the highest energies in the recorded spectrum and by considering the known position in the S2 versus S1 space. Therefore, only a coarse energy cut on S1 and S2 is used to select those events.
Figure\,\ref{fig:rn222_sel} shows the $^{222}$Rn data in this parameter space for an electric field of $ 1.24$\,kV/cm.
Three different decay lines can be identified originating from $^{222}$Rn, $^{218}$Po and $^{214}$Po at energies of 5.5, 6.0 and 7.7\,MeV, respectively. These three lines are only separable for electric fields above 70\,V/cm. For lower fields, the S2 signal size is significantly reduced due to an increased probability of electron-ion recombination in the dense tracks created by alpha particles, making the different decay lines inseparable within our S2 resolution. 
A projection of the alpha lines onto the S2 axis can be seen in figure\,\ref{fig:rn222_cs2_spectrum}. With the alpha decay energies of $^{222}$Rn and $^{218}$Po being very close, the two corresponding peaks clearly overlap. Their positions however can still be fitted separately to extract light and charge yield for both lines.
The energy resolution, defined as the ratio between the standard deviation and the mean of the fitted Gaussian, for alpha interactions is found to be around 5\% both for S1 and S2 signals at 1.2\,kV/cm. While the S1 resolution is almost constant for varying drift fields, the S2 resolution worsens slightly with decreasing electric field due to the lower charge collection. For the $^{83\textrm{m}}$Kr signals, we obtain 29\% and 20\% resolution for the 9.4 and 32.1\,keV S1 lines and  19\% for the combined 41.5\,keV S2 line.

	
\section{Response of liquid xenon to electrons and $\mathbf{\upalpha}$-particles}
\label{sec:response}

The creation of scintillation and ionization signals in LXe is a complex process which has been investigated intensively over the past decades\,\cite{Aprile:2006kx, Aprile:2007qd, Manalaysay:2009yq, Lin:2013ypa, Hogenbirk:2018knr, Temples:2021jym}. 
In the absence of an electric field, ionization electrons recombine with the positively charged ions, leading to additional S1 scintillation photons.
In the presence of an applied field, the electrons are drifted out of the interaction site to be read out separately. The amount of charge and light depend on the electric field, i.e. the higher the field, the more efficiently electrons are prevented from recombining, leading to a higher charge yield and a lower light yield. For high enough fields, the charge and light yields reach a plateau.
\vskip 0.03cm

Alpha particles have a high dE/dx and therefore the ionization cloud is significantly denser than for electrons. As a result, most electrons are screened by the surrounding ionized medium and cannot escape recombination. The light yield is consequently higher and the charge yield lower than for an electron of equivalent energy. In this section, we investigate the signal production for low energy electrons and alpha particles using the data taken with the ${}^\mathrm{83m}\mathrm{Kr}$ and $^{222}$Rn sources. 
\vskip 0.03cm

\begin{figure}[h]
    \centering
    \includegraphics[width=0.5\textwidth]{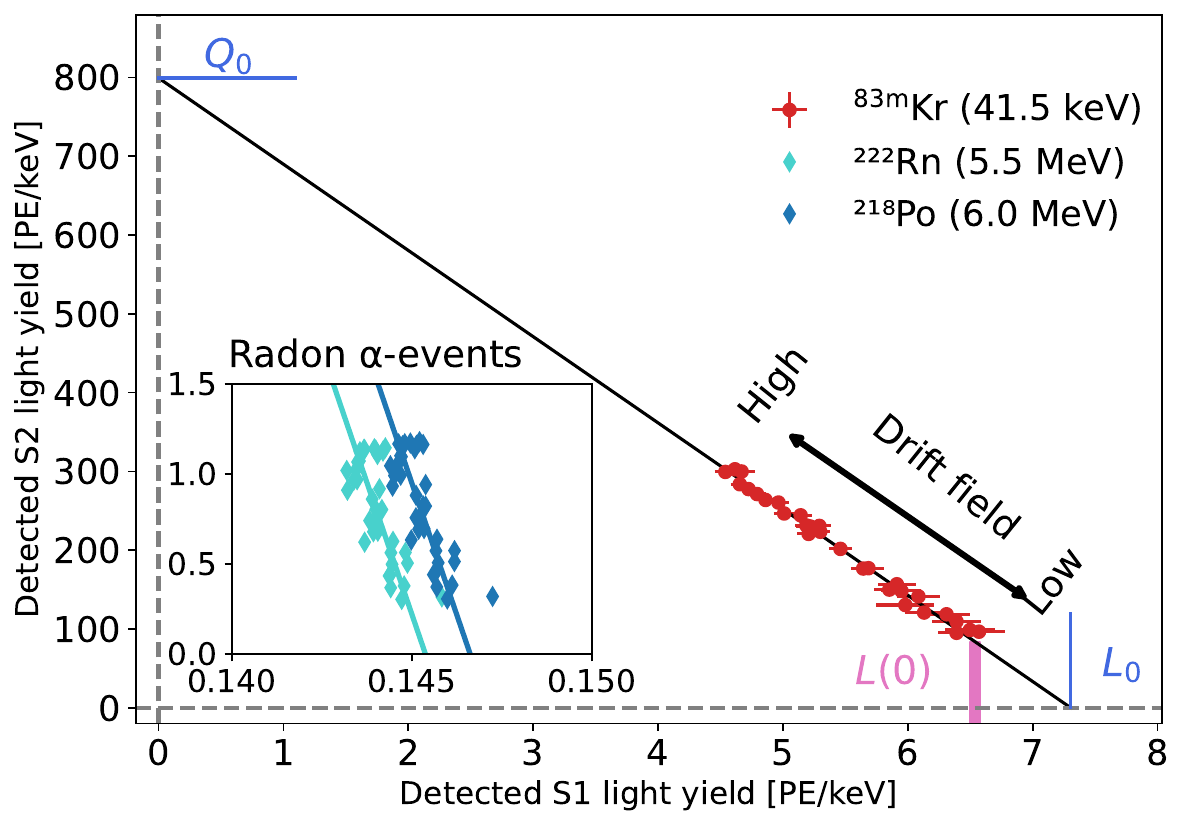}
    \caption{Anti-correlation of charge and light for the 41.5\,keV line of the $^{83\textrm{m}}$Kr source (red) and for the alpha particles of $^{222}$Rn and $^{218}$Po in the lower left inset. Only data points with a drift field above 70\,V/cm are shown as only those are used for the fit. For better visibility, the data points of $^{214}$Po are not shown in the inset.}
    \label{fig:s1_s2_anticor}
\end{figure}

First, we parameterize the S1-S2 anti-correlation in our TPC using the measurements performed at various drift fields. Figure\,\ref{fig:s1_s2_anticor} shows the corrected S1 and S2 values for different particles. 
In the case of ${}^\mathrm{83m}\mathrm{Kr}$, the light yield of the two transitions at 9.4\,keV and 32.1\,keV can be measured separately. However, due to the short half-life of the intermediate state, the S2 signals overlap and only a combined measurement (41.5\,keV) is possible for the charge yield. For this reason, the main panel of figure\,\ref{fig:s1_s2_anticor} contains only the 41.5\,keV data.
The observed anti-correlation is then fit with a linear function and the values for $L_0$ and $Q_0$ are determined as the intercepts with the axis.
The value of $L(0)$ is obtained by extrapolation of the light signal to zero field using equation\,\ref{eq:ti_funcion}.
Its value is found to be smaller by about 10\% compared to $L_0$ due to the incomplete recombination of electron-ion pairs at zero field.
We therefore estimate the recombination fraction at zero field to be $(90\pm2)$\%.
\vskip 0.02cm

\begin{figure}[h]
    \centering
    \includegraphics[width=0.499\textwidth]{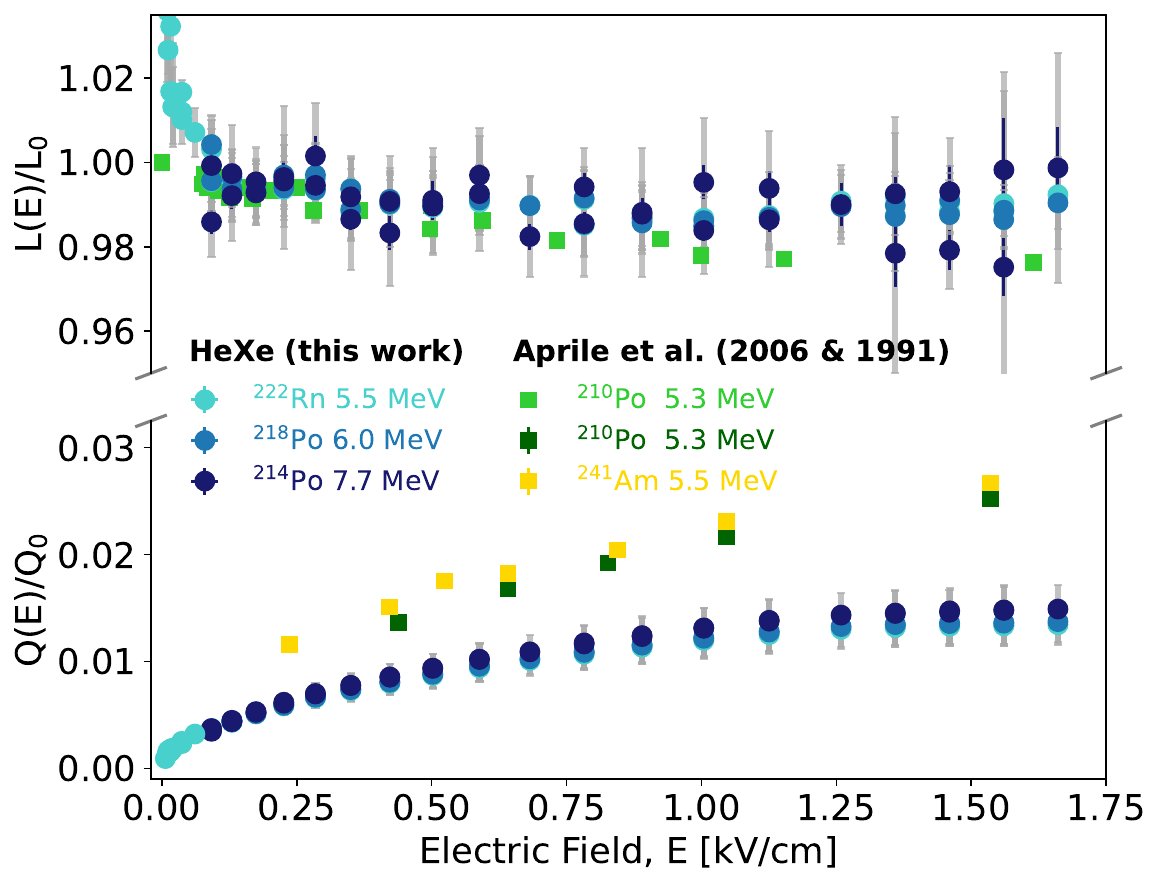}
    \caption{Light and charge yields for the $\upalpha$-decays of $^{222}$Rn, $^{218}$Po, and $^{214}$Po as function of electric field.  The data is compared with the results from~\cite{Aprile:1991xb} and \cite{Aprile:2006kx}. Gray bars indicate the systematic error. Data points are provided online\,\cite{HeXeData}.}
    \label{fig:rn_yields}
\end{figure}

The $^{222}$Rn data is shown in an inset. Note that the detected S1 and S2 signals are not comparable between the two datasets due to the presence of attenuators in the $^{222}$Rn case which modifies the light collection.  For the $^{222}$Rn data, we exclude the data points below 70\,V/cm from figure\,\ref{fig:s1_s2_anticor}, as the data quality is significantly better for drift fields above that. 
Overall data points are very clustered (due to the small quenching of alpha particles) and the fit has a significant uncertainty. This uncertainty is propagated as systematic error and it is represented in figure\,\ref{fig:rn_yields} by gray error bars. The data points for $^{222}$Rn and $^{218}$Po are shifted from each other (approx. 1\%), likely due to the different dE/dx of the two alpha lines.
\vskip 0.02cm

\begin{figure}[h]
    \centering
    \includegraphics[width=0.48\textwidth]{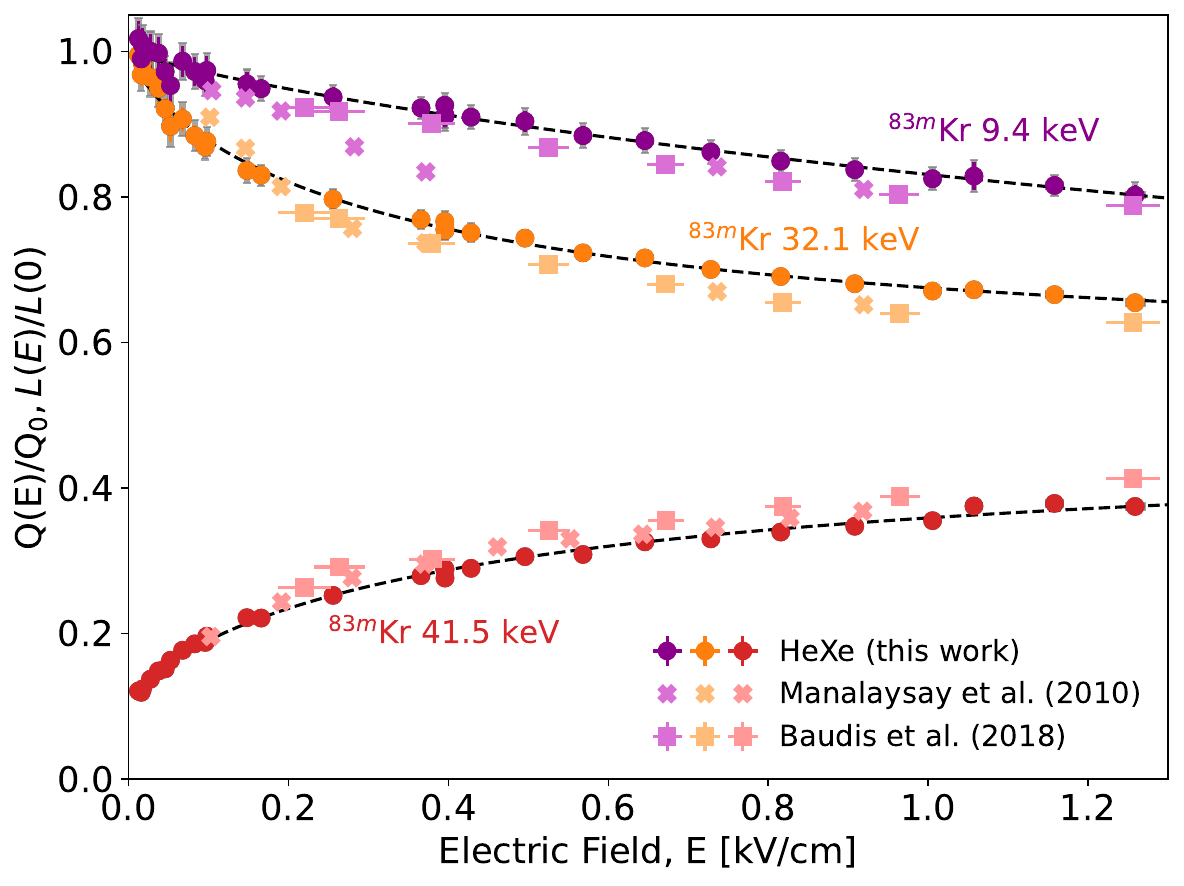}
    \caption{Light and charge yields of the ${}^\mathrm{83m}\mathrm{Kr}$ transitions as function of applied electric field $E$, measured using the HeXe setup (circles). The data points from Manalaysay et al.\,\cite{Manalaysay:2009yq} (crosses) and Baudis et al.\,\cite{Baudis:2017xov} (squares) are also shown. The dotted lines show a fit with the modified Thomas-Imel box model given in equation\,\ref{eq:ti_funcion} (data points are provided online\,\cite{HeXeData}).}
    \label{fig:kr_yields}
\end{figure}

Figure\,\ref{fig:kr_yields} shows the field dependence of the normalized charge and light signals for the ${}^\mathrm{83m}\mathrm{Kr}$ decay lines.
The combined 41.5\,keV S2 signal (red) is normalized by $Q_0$ as determined in figure\,\ref{fig:s1_s2_anticor}, while the light signals from the 9.4\,keV and 32.1\,keV decays (purple and orange) are normalized to their extrapolated values at zero field ($L(0)$).
To avoid the variation of the 9.4\,keV signal yield with the time between the 32.1\,keV and the 9.4\,keV S1 signals as seen in~\cite{Baudis:2013cca}, we select events in which the two corresponding S1 signals are more than 300\,ns apart.
We find a good agreement between our data and the phenomenological function (black dashed line) used in~\cite{Manalaysay:2009yq}, which is based on the Thomas-Imel box model\,\cite{Thomas:1987zz}, expressed as:
\begin{align}
\frac{L(E)}{L(0)}, \frac{Q(E)}{Q_{0}} = a_1\cdot a_2 \cdot E \cdot \ln{\left(1+\frac{1}{a_2\cdot E}\right)} + a_3\,.\label{eq:ti_funcion}
\end{align}
$E$ is the applied electric field, $L(E)$ and $Q(E)$ are the light and charge yields with their normalization constants $L(0)$ and $Q_0$, and $a_1$, $a_2$ and $a_3$ are fit parameters. Their results and statistical uncertainties are reported in table\,\ref{tab:kr_yields_fit_vals}.
\begin{table}[h]
	\centering
	\begin{threeparttable}
		\caption{Best fit values for the field dependent light and charge yield variation for ${}^\mathrm{83m}\mathrm{Kr}$ (shown in figure\,\ref{fig:kr_yields}) explained by the modified Thomas-Imel box model shown in equation\,\ref{eq:ti_funcion}. Only statistical uncertainties are reported.}\label{tab:kr_yields_fit_vals}
		\centering
		\begin{tabular}{r D{,}{\pm}{-1} D{,}{\pm}{-1} D{,}{\pm}{-1}}
			\toprule
		    Transition   & \multicolumn{1}{c}{$a_1$}    & \multicolumn{1}{c}{$a_2\,\mathrm{\left[cm/kV\right]}$}    &\multicolumn{1}{c}{$a_3$}    \\
			\midrule
			9.4\,keV          & -1.0\,,\,0.3                 & 0.06\,,\,0.03 &   \multicolumn{1}{r}{1 (fixed)}\\
            32.1\,keV         & -0.443\,,\,0.011                 & 1.41\,,\,0.10 &   \multicolumn{1}{r}{1 (fixed)}\\
			41.5\,keV         & 0.375\,,\,0.008                 & 1.11\,,\,0.11 &   0.091\,,\,0.004\\
			\bottomrule
		\end{tabular}
	\end{threeparttable}
\end{table}

Our data is compared to the data reported in Manalaysay et al.\,\cite{Manalaysay:2009yq} and Baudis et al.\,\cite{Baudis:2017xov}.
For the charge yield of the summed signal at 41.5\,keV, we find a similar dependence on electric field but our values are systematically lower.
Note that the charge yield data reported in\,\cite{Manalaysay:2009yq} has been scaled by 0.943 to match our definition of $Q_0$ and for the data from\,\cite{Baudis:2017xov}, a value for $Q_0$ was determined and combined with the reported value of $g_2$.
For the light yield, we find a similar good agreement for drift fields above 0.4\,kV/cm with our data points being systematically slightly above the literature measurements.
Below this value, the shape found in\,\cite{Manalaysay:2009yq} differs slightly.
The light yield data reported in\,\cite{Baudis:2017xov} were normalized by an estimated value of $L_0$ and their reported value for $g_1$ and then scaled by our observed recombination fraction at zero field.

Note that our data include several measurements at drift fields below 200\,V/cm which were never measured systematically before.
\vskip 0.03cm

\begin{figure}[h]
    \centering
    \includegraphics[width=0.48\textwidth]{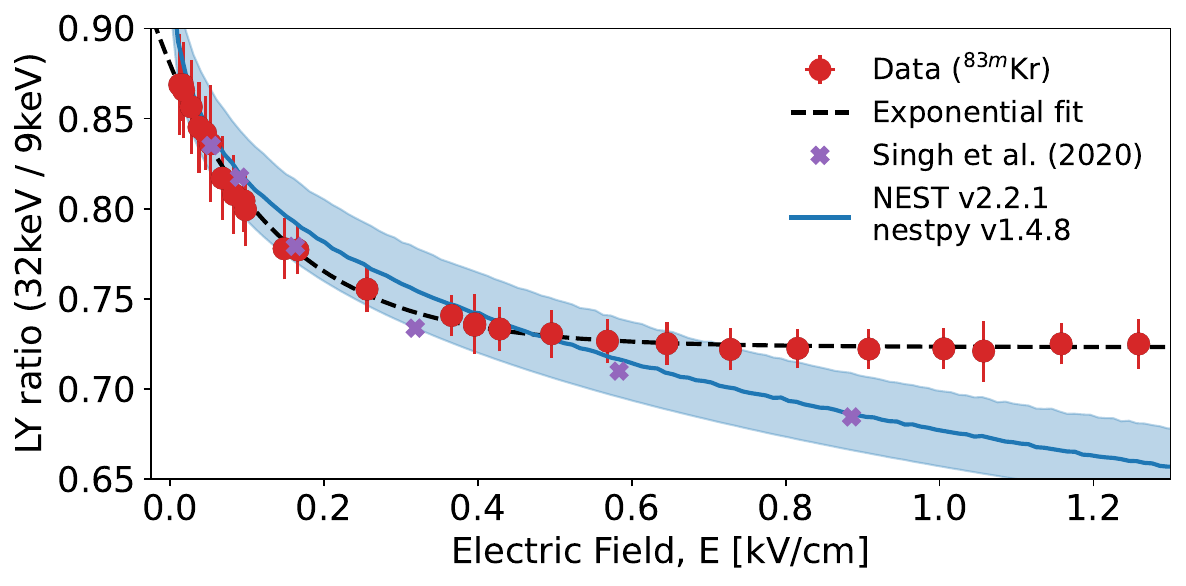}
    \caption{Dependence of the ratio between the observed light yields from both ${}^\mathrm{83m}\mathrm{Kr}$ transitions on the applied drift field (data points are provided online\,\cite{HeXeData}). A fit of the data with equation\,\ref{eq:kr_ratio_function} is shown by the dashed black line alongside with the prediction of the NEST\,\cite{Szydagis_2011,nestpy:1_4_9} simulation framework and data reported by Singh et al.\,\cite{Singh:2019nrd} (purple).}
    \label{fig:kr_yields_ratio}
\end{figure}

The ratio between the S1 light yield of both ${}^\mathrm{83m}\mathrm{Kr}$ transitions is of special interest as it is quite sensitive to the electric field.
Therefore, it can be used to map the magnitude and variation of the drift field in large-scale detectors, as done for example in\,\cite{LUX:2017fwq}.
Since both decays occur at the same physical location, the ratio is largely unaffected by geometric effects like the light collection efficiency.
Figure\,\ref{fig:kr_yields_ratio} shows the dependence of this ratio over the range of the investigated electric drift fields.
The ratio is defined as the observed light yield normalized to the decay energy (PE/keV) and its dependence is found to be well described by the following phenomenological function (dashed black line):
\begin{align}
R(E) = \mathrm{\frac{LY\,(32\,keV)}{LY\,(9\,keV)}} = b_1\cdot e^{-b_2\cdot E} + b_3\,.\label{eq:kr_ratio_function}
\end{align}
The extracted fit parameters with their statistical uncertainties are $b_1 = 0.156\pm 0.012$, $b_2 = (6.6\pm1.1)$\,cm/kV and $b_3 = 0.723 \pm 0.004$.
The prediction of the NEST simulation framework\,\cite{Szydagis_2011,nestpy:1_4_9} is shown in blue with the shaded area representing the range due to the time difference between the first and second ${}^\mathrm{83m}\mathrm{Kr}$ transition ($\Delta t > 300$\,ns).
For fields $\lesssim$ 500\,V/cm, we observe only small deviations with the ratio predicted by NEST being larger by approximately 2\%.
For fields above 500\,V/cm, however, we find that the ratio does not decrease with an increasing field which is in contradiction to the NEST model.
In the data from Singh et al.\,\cite{Singh:2019nrd} (in purple), the same requirement on the time separation of both ${}^\mathrm{83m}\mathrm{Kr}$ transitions ($\Delta t > 300$\,ns) is used.
\vskip 0.03cm

The light and charge yields of $\upalpha$-particles from the $^{222}$Rn source are measured in a new run because Teflon attenuators need to be introduced to prevent signal saturation. For fields above 70\,V/cm, we fit the data with a sum of three individual 2D Gaussian functions (see figure\,\ref{fig:rn222_sel}) obtaining light and charge yields of $^{222}$Rn, $^{218}$Po, and $^{214}$Po separately. For lower fields the $\upalpha$-decays from $^{222}$Rn and $^{218}$Po become indistinguishable and therefore, only a combined fit is possible. The $^{222}$Rn light yield values are corrected taking into account the bias produced by the fitting of two lines together. 
Figure\,\ref{fig:rn_yields} shows the resulting light and charge yields as function of electric field for alpha particles. 
The uncertainty on the field strength (too small to be visible in the figure) is calculated from the electric field simulation as the width of the central 68th percentile of the electric field values throughout the fiducial volume. 
The vertical error bars include a statistical error from the number of events as well as a systematic error (gray). The latter is estimated by varying analysis parameters like selection cuts or the fiducial volume and it is combined with the uncertainty due to the normalization method described above.
The uncertainty of the charge yield is mostly due to the normalization $Q_0$, while the uncertainty of the light yield dominated by the variation of the analysis parameters.

\vskip 0.02cm

The data is shown together with existing data from Aprile et al.\,\cite{Aprile:1991xb,Aprile:2006kx}. While the light yield measurements agree with each other within the uncertainty, it can be seen that for our measurements at fields below $\sim 70$\,V/cm the normalized light yields exceed values of 1. 
We attribute this to field inhomogeneities due to field leakage from the extraction region into the drift region at lower fields (see figure\,\ref{fig:Efield_scan}). This in turn causes the S1 light collection correction to become less effective and therefore these data points are excluded from the normalization shown in figure\,\ref{fig:s1_s2_anticor}.

To compare the charge yield dependence between both measurements, a scaling needs to be applied to the data points from\,\cite{Aprile:1991xb}, in order to reflect the same normalization method as is used for our data (see figure\,\ref{fig:s1_s2_anticor}).
The data points in~\cite{Aprile:1991xb} are multiplied by the ratio between $W=11.5$\,eV\,\cite{EXO-200:2019bbx,Baudis:2021dsq} which represents the mean energy to produce a quantum in the LXe and $W_i = 15.6$\,eV, which gives the mean energy required to produce an electron ion pair.
The curve of the charge yield is found to be significantly lower as compared to the values reported in\,\cite{Aprile:1991xb, Aprile:2006kx} and flattens slightly above 1.3\,kV/cm.
We discuss in the following possible explanations for the discrepancy based on both different experimental conditions and the different data analysis.
\vskip 0.02cm

The reported liquid xenon temperature in\,\cite{Aprile:1991xb} is 195\,K, which is approximately 20\,K higher compared to our measurement. Consequently, the liquid xenon density in our measurement was higher by about 5.1\%\,\cite{AirLiquid}. This should lead to a slightly higher ionization density as well, resulting in a lower charge yield.
Moreover, while we use a homogeneously distributed source, for the measurements reported in~\cite{Aprile:1991xb,Aprile:2006kx} the source was directly deposited onto the cathode surface. 
This restricts the detectable alpha emission angles to the 2$\uppi$ upward direction, whereas in our measurement alpha particles emitted under all possible angles are detectable.
Horizontal $\upalpha$-tracks are expected to have a slightly higher charge yield compared to tracks which are aligned with the field (see also discussion in\,\cite{Aprile:1991xb}). For vertically aligned tracks, electrons need to drift through the volume of the track, where a high ion-density is present, increasing their probability to recombine. Our measurement contains all 4$\uppi$ possible directions of the alpha particle which could also contribute to the observed difference.

Furthermore, a small fraction of the alpha decay energy is carried away by the recoil of the daughter nucleus. 
In our measurement, we observe the sum of the alpha-particle energy and the $\mathcal{O} (100\,\textrm{keV})$ recoil of the daughter nucleus, while in~\cite{Aprile:1991xb,Aprile:2006kx} this recoil is absorbed by the cathode and therefore not detectable.
In our data, we can make use of a fiducial volume selection, to remove events close to the electrodes avoiding the strongly distorted fields in those regions. Using our detailed 3D detector simulation shown in section\,\ref{sec:efield}, we have a good knowledge of the electric field and its uncertainty within our active volume.
The field strength close to the cathode surface in\,\cite{Aprile:1991xb,Aprile:2006kx} might be enhanced due to surface irregularities. This possibility was considered and mitigated by polishing the cathode prior to the deposition of the sources. Nevertheless, this could contribute slightly to the discrepancy.
\vskip 0.02cm


\section{Drift velocity measurement}
\label{sec:drift}
	
In liquid xenon TPCs, the $z$-coordinate or depth of an event is reconstructed from the time difference between the prompt scintillation light (S1) and the proportional light from the drifted electrons (S2). The conversion from time to distance can be easily performed using the drift velocity which is usually constant when the electric field is sufficiently homogeneous. This section describes a measurement of the electron drift velocity in liquid xenon at a temperature of $(174.4 \pm 0.2)$\,K for electric field values between 7.5 and 1\,642\,V/cm.
\vskip 0.02cm

\begin{figure}[h]
    \centering
    \includegraphics[width=0.49\textwidth]{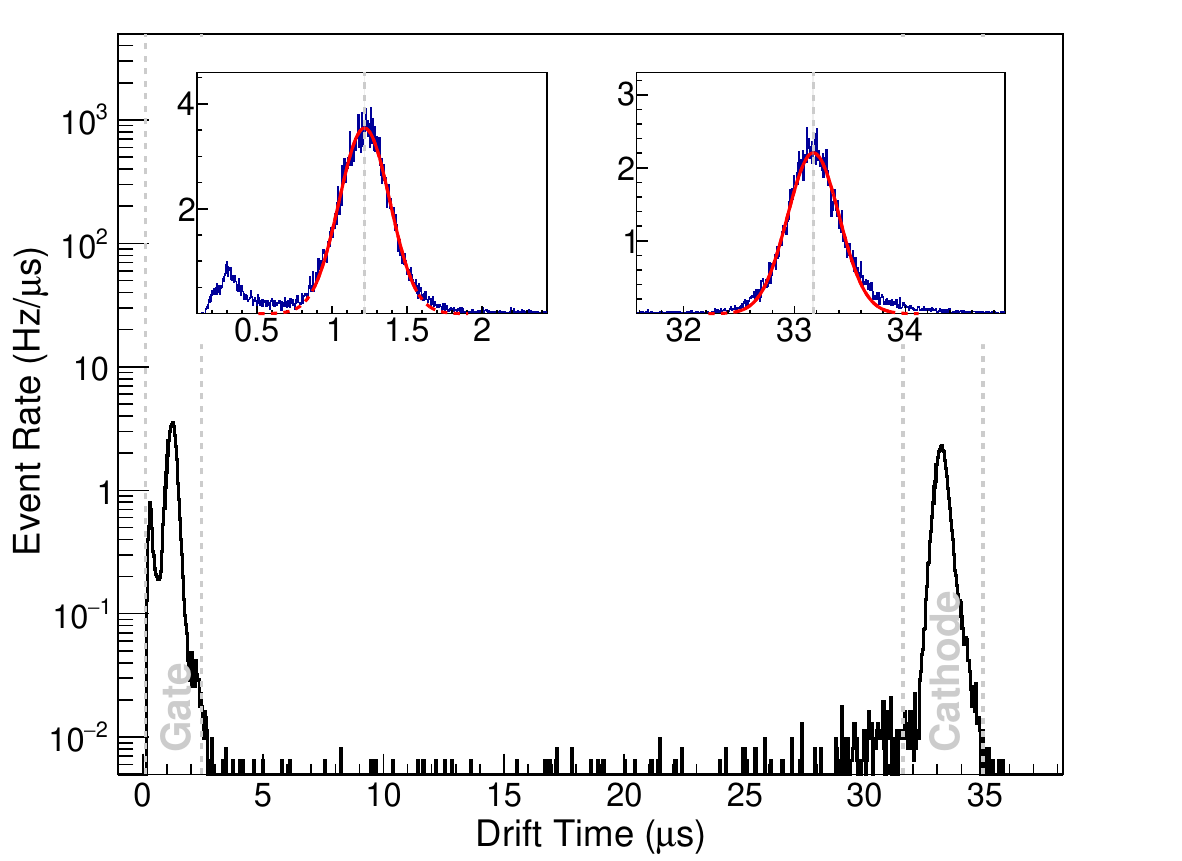}
    \caption{Drift time distribution of small-sized S2 events for $^{222}$Rn data in the TPC for a field of 225\,V/cm zooming into the gate region (left inset) and the cathode region (right inset).}
    \label{fig:GateCathode}
\end{figure}

To extract the drift velocity, we divide the distance between gate and cathode, which amounts to $\mathrm{(49.3 \pm 0.3)\,mm}$ at the liquid xenon temperature, by the time that the electrons take to travel between these electrodes. The uncertainty of this drift length takes machining tolerances and the uncertainty related to the thermal contraction of the PTFE into account.
The time is extracted from the features that alpha-particle energy depositions produce at the gate and the cathode. Since alpha particles produce very bright S1 signals, electrons are released from the stainless steel surfaces of the meshes via the photoelectric effect. This happens in time coincidence with the S1 signal. The emitted electrons are drifted to the amplification region, where they produce small S2 signals. Due to the fixed drift distances, the positions of the electrodes can be recognized in the data as peaked structures. Figure\,\ref{fig:GateCathode} shows the drift time distribution of events with S2 signals smaller than 25\,PE zooming into the gate region (left inset) and the cathode region (right inset).
The position of the gate is determined by fitting a Gaussian function to the peak. The cathode position is fit using a Gaussian function for fields larger than 17\,V/cm and a right-tailed Crystal-ball function for lower drift-fields (red lines in figure\,\ref{fig:GateCathode}).
For each drift-field configuration, both fits are repeated 20 times, where the fit region is varied within $[\pm 1\sigma, \ldots \pm 2\sigma]$.
The maximum variation between the fitted mean positions is then added to the uncertainty from the geometric distance.
\vskip 0.02cm

The drift velocity of electrons can also be determined using the $^{83\textrm{m}}$Kr  data acquired at an operational temperature of $(173.41\pm 0.11)$\,K in a drift field range between 13.5\,V/cm to 1\,245\,V/cm. In this case, the positions of gate and cathode are determined in a different way. The charge yield in the extraction region typically differs from the one in the drift volume. The change in the S2 size of $^{83\textrm{m}}$Kr  can be therefore used to determine the position of the gate. 
The position of the cathode coincides with a steep decrease of the event rate, since no electrons can be drifted upwards from the region below the cathode. We determined the cathode position as the time where this decrease is maximal, following a similar approach as reported in\,\cite{Baudis:2017xov}.
\vskip 0.02cm
	
\begin{figure}[h]
	\centering
	\includegraphics[width=0.49\textwidth]{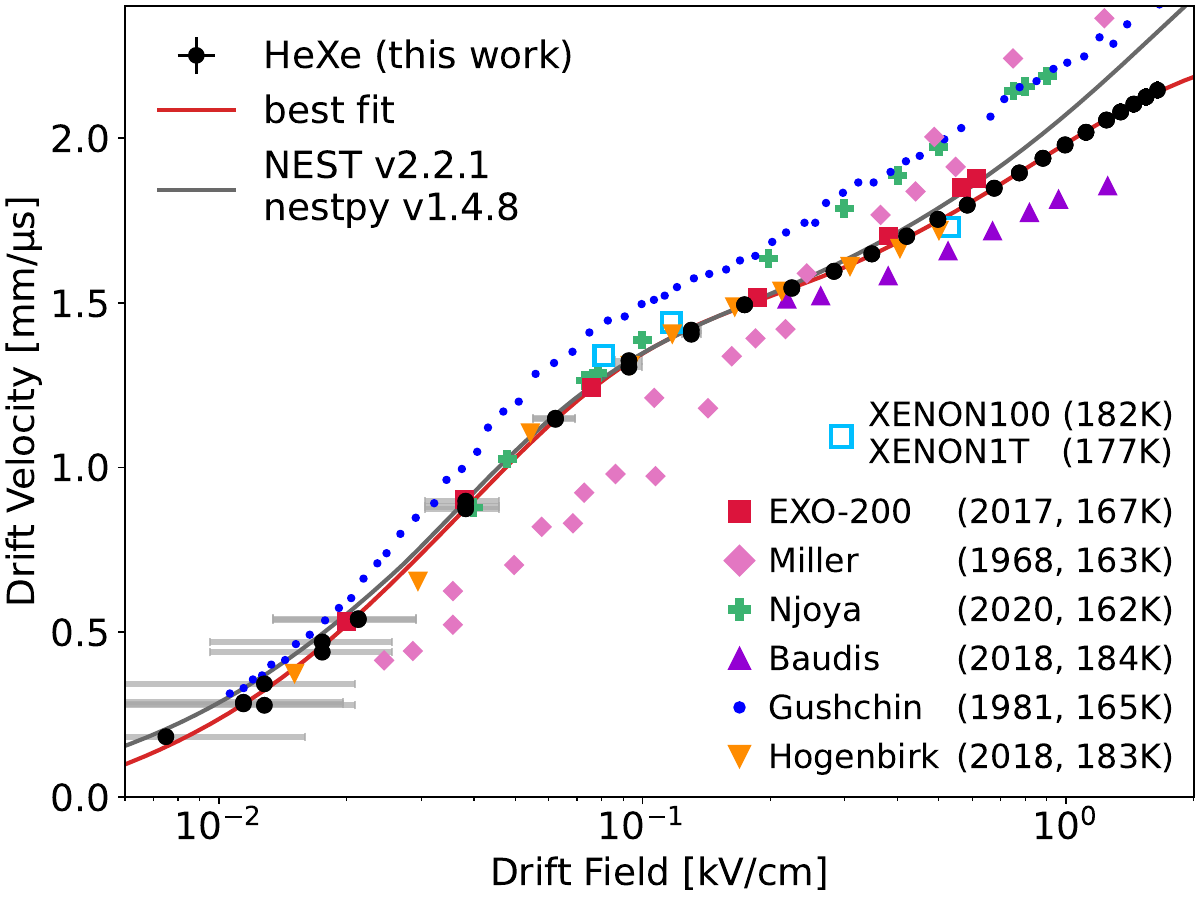}
	\caption{Drift velocity of electrons in LXe from alpha-particle interactions as measured using the HeXe setup at T$=174\,\mathrm{K}$ as a function of applied drift field. Horizontal gray bars indicate the central 68\% quantile range of the field values throughout the drift volume. For better visibility the data points from $^{83\textrm{m}}$Kr are not shown but are provided online\,\cite{HeXeData}.}
	\label{fig:drift_velocity}
\end{figure}

The derived drift velocity values from the alpha data are displayed in figure \ref{fig:drift_velocity}. 
For the x-axis, the vertical component of the electric field is taken as this is what determines the electron drift. For most of the data, this vertical component corresponds to the absolute value of the electric field, except for very low fields for which the absolute field is higher due to field distortions.
The gray horizontal bars represent the range of the field deviation throughout the active volume.
Most data points were measured twice, however, they are not separately visible in the figure because they lie exactly on top of each other. The only exception is the point at 13\,V/cm, where in one of the two datasets the data quality was significantly poorer for unknown reasons, likely due to charge up following a HV trip.
Data from the EXO-200\,\cite{Albert:2016bhh}, XENON100\,\cite{Aprile:2011dd} and XENON1T\,\cite{XENON:2019izt} experiments are displayed together with the measurements of dedicated setups\,\cite{Baudis:2017xov, Hogenbirk:2018knr, Miller:1968zza, Gushchin:1982b, Njoya:2019ldm}. The drift velocity derived using $^{83\textrm{m}}$Kr  events are in agreement with the ones in the figure within errors.
\vskip 0.02cm

\begin{figure}[h]
	\centering
	\includegraphics[width=0.5\textwidth]{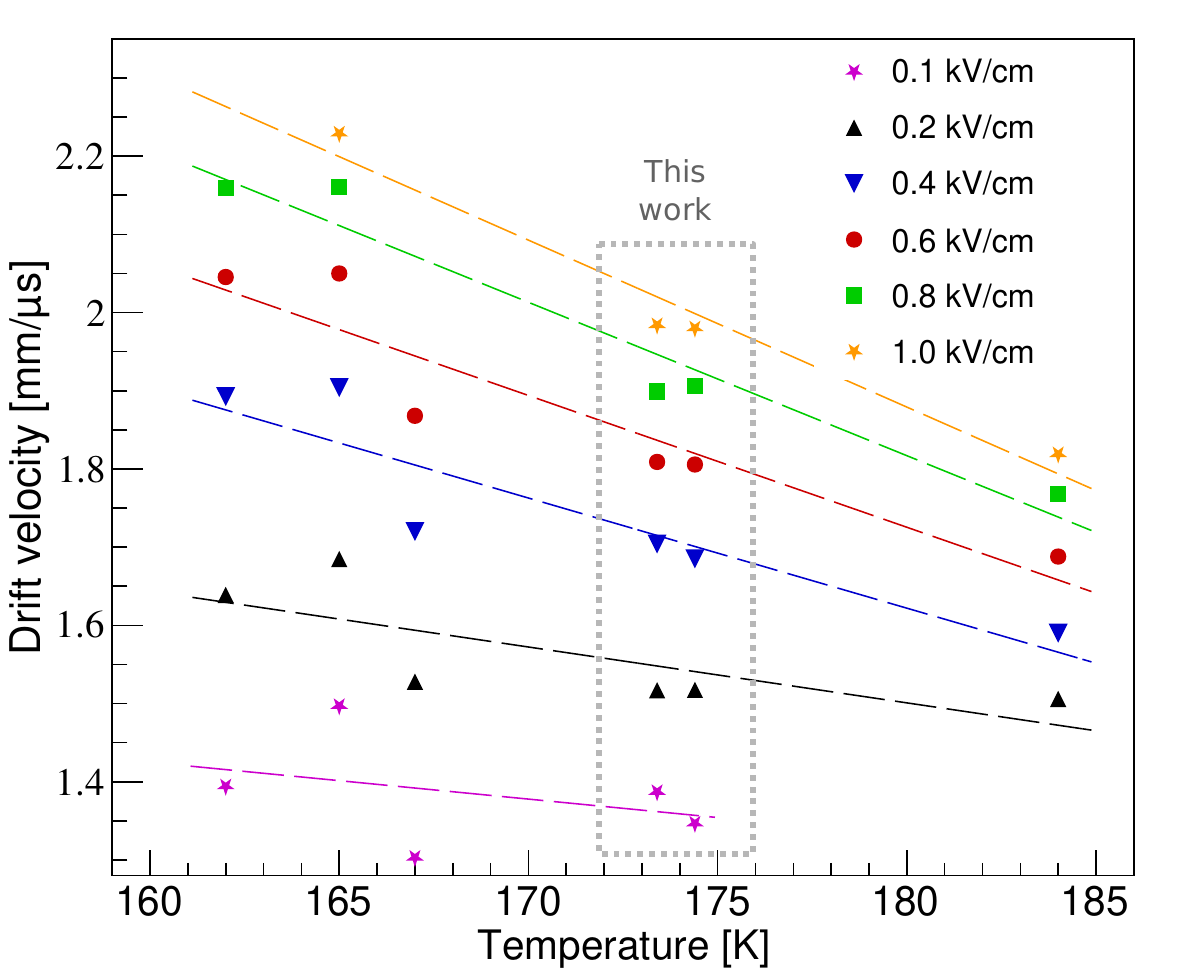}
	\caption{Dependence of the drift velocity on the temperature for a few selected values of the drift field.}
	\label{fig:drift_T}
\end{figure}

Our data covers a large range in electric fields from 7.5 up to 1\,642\,V/cm and is overall in good agreement with literature data. The systematic discrepancies with some of the data sets are attributed to the different operating temperatures as argued in~\cite{Baudis:2017xov}. The data from Baudis for instance shows lower drift velocity values due to the higher operating temperature of 184\,K. 
Following the same argument, the drift velocity values from Miller and EXO are larger as they were acquired at 163\,K and 167\,K, respectively (see also figure\,\ref{fig:drift_T}).
\vskip 0.02cm

We fit the data with a model based on the phenomenological approach of the NEST\,\cite{Szydagis_2011} framework using a sum of individual exponential functions and an offset to describe the dependence of drift velocity $v_{d}$ to the applied drift field $E$.
We find that the data within the investigated range of drift fields can be sufficiently described using only two exponential functions:
\begin{align}
v_{d}(E) = A_1 \cdot e^{-E/B_1} + A_2 \cdot e^{-E/B_2} + C\,.\label{eq:v_drift_fit}
\end{align}
The constants $A_{1,2}$, $B_{1,2}$ and $C$ are determined from data for a given LXe temperature. Table~\ref{tab:fit_drift_velocity} summarizes the found values for the two calibration sources $^{83\textrm{m}}$Kr as well as $^{222}$Rn.
The comparison between the measured data and the prediction of the NEST framework\,\cite{Szydagis_2011,nestpy:1_4_9} (gray line) shows reasonable agreement in the low field regime, while for higher field we observe systematically smaller drift velocities.
\begin{table}[h]
	\centering
	\begin{threeparttable}
		\caption{Summary of fit parameters to equation~\ref{eq:v_drift_fit} for $^{83\textrm{m}}$Kr  and $^{222}$Rn measurements at given temperatures}\label{tab:fit_drift_velocity}
		\centering
        \begin{tabular}{l D{,}{\pm}{-1} D{,}{\pm}{-1} c}
			\toprule
		    Parameter   & \multicolumn{1}{c}{$^{83\textrm{m}}$Kr}    & \multicolumn{1}{c}{$^{222}$Rn}   &  Unit    \\
			\midrule
			Temperature & 173.41\,,\,0.11                 & 174.4\,,\,0.2 &   K\\
			$A_1$       & -1.38\,,\,0.02                  & -1.458\,,\,0.013    &  \\
			$A_2$       & -0.95\,,\,0.13                  & -0.95\,,\,0.02      &  \multirow{-2}{*}{$\mathrm{mm/\upmu s}$}\\
			$B_1$       & 38.0\,,\,1.8                   & 34.7\,,\,1.2        &  \\
			$B_2$       & 1000\,,\,300                    & 830\,,\,70          &  \multirow{-2}{*}{$\mathrm{V/cm}$}\\
			$C$         & 2.33\,,\,0.15                   & 2.27\,,\,0.04       &  $\mathrm{mm/\upmu s}$ \\
			\bottomrule
		\end{tabular}
	\end{threeparttable}
\end{table}
\vskip 0.02cm

The dependence of the drift velocity on the LXe temperature is shown in figure\,\ref{fig:drift_T} for a range of drift fields between 0.1 and 1\,kV/cm where several measurements exist.
A subset of data sets from figure\,\ref{fig:drift_velocity} is employed for which the measurement temperature was clearly stated. Data at 162\,K~\cite{Njoya:2019ldm}, 165\,K~\cite{Gushchin:1982b}, 167\,K~\cite{Albert:2016bhh}, 
173.4\,K and 174.4\,K (this study) and 184\,K~\cite{Baudis:2017xov} is used. 
The values for the drift velocity at each electric field are determined interpolating linearly in between the published data when necessary. 
The electron mobility depends on temperature roughly as $\mu_e \propto T^{-3/2}$~\cite{Miller:1968zza}. Using the dependence of the drift velocity on the electric field, $v_d = \mu \cdot E$, the data of figure\,\ref{fig:drift_velocity} could in principle be fit with a $\propto T^{-3/2}$ dependence. However, this functional dependence is unable to describe the steep increase of drift velocity at cold temperatures for high fields ($E>0.6$\,kV/cm).
Therefore and just to guide the eye, we fit a linear function to the points of each electric field. 
Our data points are acquired at a temperature close to the expected operating temperature of XENONnT and therefore are of interest for the experiment.
\vskip 0.02cm

\section{Conclusions}
\label{sec:conclusion}

In this work, we describe the performance of our small-scale liquid-xenon TPC, HeXe, which is a multipurpose R\&D detector. A new 3 dimensional electric field simulation using COMSOL Multiphysics\textsuperscript{\tiny\textregistered} allows us to optimize the design for a uniform electric field and to quantify the error on the field which is often ignored. This is especially important for the measurements performed at fields below $\sim 100$\,V/cm. 

We employ data from a $^{83\textrm{m}}$Kr and a $^{222}$Rn source to characterize the light and charge yields of the corresponding decay lines. 
For $^{83\textrm{m}}$Kr, we systematically investigate low drift fields down to 13.5 V/cm where no measurements existed in the literature so far. We study the signal yields of the three $\upalpha$-lines $^{222}$Rn, $^{218}$Po and $^{214}$Po. We find a significant deviation of the charge yield compared to existing data. Although the origin of the discrepancy is not understood, we give several hypotheses including different analysis methods but also different experimental conditions.

The drift velocity of ionization electrons is measured for field values between 7.5 and 1\,640\,V/cm at a LXe temperature of 174.4\,K. Several measurements at fields below 100\,V/cm were taken to characterize the low field region for which literature is scarce. We study the temperature dependence of the drift velocity using also literature values and find an increase in drift velocity for decreasing temperatures which is most pronounced at high fields.

\begin{acknowledgements}
We acknowledge the support of the the Max Planck Society. We thank our technicians Jonas Westermann, Steffen Form,
and Michael Rei\ss felder for their technical support during construction and operation of the system. Furthermore, we thank our colleges Hardy Simgen, Stefan Br\"unner and Sebastian Lindemann for their contributions and very fruitful discussions. 

\end{acknowledgements}

\bibliographystyle{utphys}
\bibliography{manuscript_v2}   

\end{document}